\documentclass[twocolumn]{aastex631}

\usepackage{hyperref}
\usepackage{natbib}
\usepackage{graphicx}
\usepackage{epsfig}
\usepackage{amsmath}
\usepackage[english]{babel}
\shorttitle{Stellar bars in jellyfish galaxies}
\shortauthors{ S\'anchez-Garc\'ia et al.}

\begin{document}

\title{GASP XLV: Stellar Bars In Jellyfish Galaxies. Analysis of ionised gas and stellar populations}
\correspondingauthor{Osbaldo S\'anchez-Garc\'ia}
\email{o.sanchez@irya.unam.mx}

\author[0000-0002-2808-1223]{Osbaldo S\'anchez-Garc\'ia} 
\affiliation{Instituto de Radioastronom\'ia y Astrof\'isica, Universidad Nacional Aut\'onoma de M\'exico, Antigua Carretera a P\'atzcuaro \# 8701, Ex-Hda. San Jos\'e de la Huerta, Morelia, Michoac\'an, M\'exico C.P. 58089}
\author[0000-0002-2897-9121]{Bernardo Cervantes Sodi} 
\affiliation{Instituto de Radioastronom\'ia y Astrof\'isica, Universidad Nacional Aut\'onoma de M\'exico, Antigua Carretera a P\'atzcuaro \# 8701, Ex-Hda. San Jos\'e de la Huerta, Morelia, Michoac\'an, M\'exico C.P. 58089}
\author[0000-0002-7042-1965]{Jacopo Fritz} 
\affiliation{Instituto de Radioastronom\'ia y Astrof\'isica, Universidad Nacional Aut\'onoma de M\'exico, Antigua Carretera a P\'atzcuaro \# 8701, Ex-Hda. San Jos\'e de la Huerta, Morelia, Michoac\'an, M\'exico C.P. 58089}
\author[0000-0002-1688-482X]{Alessia Moretti} 
\affiliation{INAF - Astronomical Observatory of Padova, vicolo dell'Osservatorio 5, IT-35122 Padova, Italy}
\author[0000-0001-8751-8360]{Bianca M. Poggianti} 
\affiliation{INAF - Astronomical Observatory of Padova, vicolo dell'Osservatorio 5, IT-35122 Padova, Italy}
\author[0000-0002-1734-8455]{Koshy George} 
\affiliation{Faculty of Physics, Ludwig-Maximilians-Universit{\"a}t, Scheinerstr. 1, Munich, 81679, Germany}
\author[0000-0002-7296-9780]{Marco Gullieuszik} 
\affiliation{INAF - Astronomical Observatory of Padova, vicolo dell'Osservatorio 5, IT-35122 Padova, Italy}
\author[0000-0003-0980-1499]{Benedetta Vulcani} 
\affiliation{INAF - Astronomical Observatory of Padova, vicolo dell'Osservatorio 5, IT-35122 Padova, Italy}
\author[0000-0002-5660-1054]{Giovanni Fasano} 
\affiliation{Independent Researcher}
\author{Amira A. Tawfeek}
\affiliation{Instituto de Radioastronom\'ia y Astrof\'isica, Universidad Nacional Aut\'onoma de M\'exico, Antigua Carretera a P\'atzcuaro \# 8701, Ex-Hda. San Jos\'e de la Huerta, Morelia, Michoac\'an, M\'exico C.P. 58089}
\affiliation{National Research Institute of Astronomy and Geophysics (NRIAG), 11421 Helwan, Cairo, Egypt}

\begin{abstract}
Stellar bars have been found to substantially influence the stellar populations properties in galaxies, affecting their ability of forming stars. While this can be easily seen when studying galaxies in relatively isolated environments, such kind of analysis takes a higher degree of complexity when cluster galaxies are considered, due to the variety of interactions which can potentially occur in these denser environments. We use IFU MUSE data from the GASP survey to study the combined effect of the presence of a stellar bar and of ram pressure, on spatially resolved properties of stellar populations. We have analyzed spatially resolved indicators of both recent SFR and average stellar population ages to check for signatures of anomalous central SF activity, also taking into account for the possible presence of nuclear activity. We found an increase of central SFR in ram pressure affected galaxies when compared with unperturbed ones. The most extreme cases of increase SFR and central rejuvenation occur in barred galaxies that are at advanced stages of ram pressure stripping. For low-mass barred galaxies affected by ram pressure, the combined effect is a systematic  enhancement of the star formation activity as opposed to the case of high-mass galaxies that present both enhancement and suppression. Barred galaxies that present a suppression of their star formation activity also present signatures of nuclear activity. Our results indicate that the combined effect of the presence of a bar and a strong perturbation by ram pressure is able to trigger the central SF activity and probably ignite nuclear activity.
\end{abstract}


\section{Introduction}
The evolution of the structure, stellar populations and overall morphology of galaxies is driven by external and/or internal processes. Among the external processes that can play a preponderant role on galaxy evolution are gravitational interactions and mergers with nearby galaxies  \citep{Toomre77, Mihos94, Park09, Lin10, Peng10}, especially for the case of satellite galaxies \citep{Peng12, Wang18}, and the many kinds of interactions that galaxies experience in the group and cluster environment, such as harassment \citep{Moore96}, starvation \citep{Larson80, Balogh00}, ram pressure stripping \citep{GunnGott72}, thermal evaporation \citep{Cowie77}, and gravitational interactions with the clusters as a whole \citep{Byrd90, Valluri93}. Among those that do not directly affect the stellar component is ram pressure stripping, which can partially or completely remove gas from the interstellar medium due to the interaction between the galaxy's interstellar medium and the intracluster medium (ICM). Extreme examples of galaxies from which gas is removed by this mechanism are the so-called jellyfish galaxies, thus dubbed for the first time by \cite{Smith+10}, which usually present unilaterally stripped material in the form of filaments and knots that leave the main body of the galaxy. These types of galaxies can be found mainly in clusters at low \citep{Merluzzi+16,Poggianti17}, at intermediate \citep{Moretti22} and at high redshift \citep{McPartland+16,Boselli19}. Tails of ionized and atomic gas can easily exceed 50 kpc in length \citep[e.g.][]{Poggianti17,deb20}, so the study of this type of galaxies cannot restrict to the main body of the galaxy, but must also include outer regions of the galaxy where these tails can be detected.

Early simulations by \cite{Bekki03} showed that the high pressure exerted by the hot intracluster medium on molecular clouds, can trigger episodes of star formation in short timescales of some hundred Myr. \cite{Kronberger08} and \cite{Kapferer09} reported that the star formation can be boosted to 2$-$10 times when compared with unperturbed systems, both in the compressed central region of the galaxy and the filaments produced by the stripped gas. On the other hand,  extreme stripping leads to full quenching and significant reddening \citep{Steinhauser16}, depending on the mass of the galaxy and the particular configuration of the orbit in the given group or cluster \citep{Bekki14}. From an observational perspective, \cite{Poggianti+16}, working with a sample of local jellyfish galaxies, found that stripping candidates tend to be located above the general population of galaxies in the star formation rate$-$stellar mass relation, indicating a star formation rate excess, and \cite{Vulcani18} identified enhanced star formation in the disks of the ram-pressure affected galaxies, additional to the star formation taking place in the tails (see also \citealt{Lee22} for a dependence of the star formation activity on the host-cluster properties). These results indicate that the ram pressure exerted by the hot intracluster medium is an effective mechanism to trigger star formation in perturbed disk galaxies, a mechanism that needs to be taken into consideration when exploring internal processes that regulate the star formation activity.

Turning to internal processes, one of the most important promoters of secular evolution are stellar bars, given that they are a prevalent structure in disk galaxies in the local Universe \citep{deVaucouleurs91, Buta10, Nair10, CervantesSodi15}. Theoretical models have demonstrated that bars can form as a result of instability in self-gravitating axisymmetric disks \citep[see][for a review on the topic]{Athanassoula13} or via gravitational interactions with perturbers of different masses \citep{Noguchi96, Miwa98}. Stellar orbits in a disk galaxy become unstable and deviate from a circular path. The tiny elongations in the stars orbits grow and they get locked into place, thus developing a bar. Bars look like elongated structures crossing the center of a disc galaxy that can be characterized by their length, strength and speed pattern. In the local Universe these bars are common and depending on the bar classification method and the wavelengths at which the galaxies are observed, the bar fraction can vary from $\sim 30$\%  \citep[optical case, see][]{Masters11, Lee19} to $\sim65$\% \citep[infrared case, see][]{Eskridge00,Menendez07}. This fraction depends on several integrated properties such as: galaxy stellar mass, color, and gas fraction \citep{Masters11, CervantesSodi17,Cervantes-Sanchez17}. The fraction increases with increasing stellar mass and is lower for bluer colors  and high gas fractions. 

Stellar bars greatly influence evolution of the gas content in disks: their non-axisymetric potential makes them highly efficient in redistributing mass and angular momentum between the components of the galaxies \citep{Weinberg85,Athanassoula92, Sellwood93,Kormendy04}. The dynamical effects of the bar potential on the stellar and gas components can lead to heating of the stellar component in the central parts of the galaxies \citep{Berentzen98, Athanassoula05, Berentzen07}, inflows of gas from the outer to the internal regions \citep{Combes85, Combes93, Kubryk15}, and potentially trigger star formation by the accumulation of gas \citep{Hunt08, Coelho11}. Indeed, several observational studies found that, compared to non-barred galaxies, barred galaxies show a higher central SF activity \citep{Ellison+11, Wang12, Consolandi+17,Catalan+17}, although this effect could also be promoted by the interaction of the galaxy with a nearby companion \citep{ Ellison+11, Wang12}. In the case of galaxies with strong bars, in addition to finding an enhancement in star formation, it has also been found that it can suppress star formation \citep{Wang12, George19, Newnham19, Geron21}, which can be explained if the onset and quenching of star formation at the center of bars occurs periodically, regulated by a balance between the inflow rate and the central concentration of mass \citep{Krumholz15}. Bar-induced star formation exhausts the infalling gas quickly, helping to build-up a central bulge, that can in turn stabilize the disk and stop further gas falling in \citep{Bournaud02, Athanassoula03}. On the other hand, if the gas inflow prompted by the bar potential is stimulating star formation in the center of the galaxies, and if galaxies use their atomic gas reserves to produce stars \citep{Saintonge11}, this might explain the trend between strongly barred galaxies and their global {\sc Hi} content, with the fraction of barred galaxies decreasing with increasing the content of {\sc Hi} \citep{Masters12, CervantesSodi17}.

Regarding the global star formation rate (SFR) of barred galaxies, \cite{Kim17} reported that the star formation activity as probed by different diagnostics was, on average, lower for strongly barred galaxies than for their non-barred counterparts, similar to the results of previous works dealing with integrated properties of large samples of galaxies \citep{CervantesSodi17, Bitsakis19}. They also indicated that although the present star formation activity of barred galaxies is lower than that of non-barred ones, no significant difference is found with respect to their stellar populations. These results are derived from integrated properties of galaxies, and indicate that the influence of the bar in promoting or quenching star formation might be very local. On the other hand, in a study of galaxy clusters segregated into interacting and non-interacting clusters, \cite{Yoon20} found that the enhancement of the fraction in star forming galaxies of moderate-mass disk-dominated galaxies in interacting clusters is mostly due to the presence of the stellar bar.

More recently, \cite{Lin17}, \cite{Chown19} and \cite{Lin20}, investigating galaxies  using data taken with Integral Field Unit surveys, found that a significant number of ``turnover" galaxies, galaxies that experienced star formation in their central region in the past 1$-$2 Gyr, were barred galaxies, a strong evidence that the bar in these galaxies may be driving gas from the disk inward to trigger star formation. Furthermore, \cite{Chown19} found that the level of enhanced central star formation correlates positively with the concentration of molecular gas for the barred galaxies and \cite{Lin17} and \cite{Lin20} found a positive relationship between the radius where the turnover occurs with the radius of the bar. Thus, these results support a close link between increased central star formation and the presence of bars. These works also found that the effect of enhanced central star formation can be trigger in pair-galaxies, indicating that the presence of a bar is a sufficient, but not a necessary condition for the turnover feature to occur.

In this work we present a study of the combined effect of ram-pressure and the presence of a bar in the stellar population of a carefully selected sample of jellyfish galaxies, with the aim to investigate the concurrent outcome of the interplay of a global external environmental process with a localized internal one. Our ultimate goal is to learn if the star formation is prone to enhancement or quenching as a result of gas redistribution, removal and/or compression by the interaction with the hot intracluster medium, at the same time of being redistributed and potentially funnel inwards by the action of the stellar bar. Having galaxies at different stages of disturbance by ram-pressure, we will also investigate if the star formation activity, already affected by the presence of a stellar bar, is regulated by the degree of the external perturbation and the presence and size of the bar. The paper is organized as follows: in Section \ref{sec:data}, we present the GASP sample used in this work and in Section \ref{sec:bar}, we describe the process to identify and characterize the barred galaxies in our sample. Section \ref{sec:SINOPSIS} introduces the spectral fitting code used to derive the physical parameters analyzed in Section \ref{sec:Results} where we present the main finding of our work. Finally, a discussion of the results and the general conclusions are summarized in Section \ref{sec:conclusions}.

Throughout this work we assume a standard $\Lambda$CDM cosmological model, with $H_0 = 70\; km\; s^{-1}\; Mpc^{-1}$, $\Omega_{m} = 0.3$ and $\Omega_{\lambda} = 0.7$), as customary for GASP papers.

\begin{figure*}
  \begin{center}
    \includegraphics [width=1\hsize]{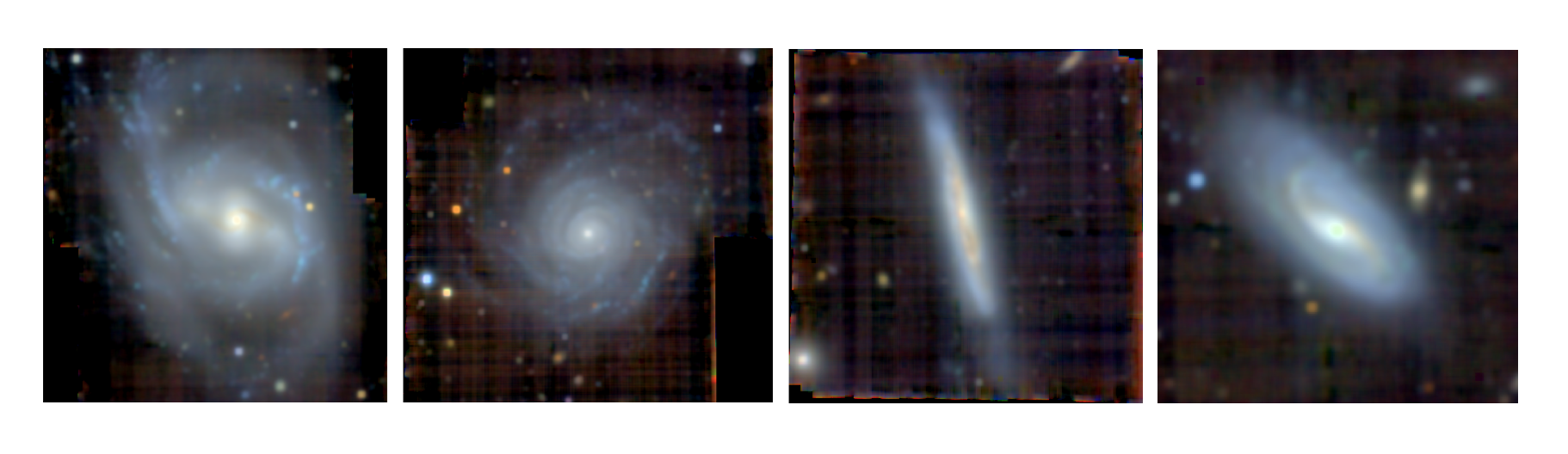}   
    \caption{\textit{From left to right:} combined $r+i+z$ color images of galaxies JO194 (barred), JO93 (unbarred), JO36 (highly inclined), JO135 (ambiguous).}
  \label{galaxies}
  \end{center}
\end{figure*}

\section{The GASP Survey}\label{sec:data}

GAs Stripping Phenomena in galaxies with MUSE \citep[GASP;][]{Poggianti17} is an integral-field spectroscopic survey with MUSE at the VLT focused at studying gas removal processes in galaxies. The GASP sample consists of 114 disk galaxies (of which 94 are primary targets and 20 compose a control sample) in the local Universe (z = 0.04 - 0.07). The field of view covered by each observation is $1' \times 1'$, which allows to observe both the galaxy disk and its outskirts, covering from 3 to 15 effective radii ($R_e$) from their center, with a mean of 7$R_e$. The galaxies in the sample are selected to probe a wide range of stellar masses, from $10^{8.7}$ to $10^{11}$ M$_{\odot}$.

The sample includes an ample range of environmental conditions since the targets observed are selected from three surveys: WINGS \citep{fasano06}, OmegaWINGS \citep{gullieuszik15} and the PM2GC \citep{calvi11}, where the first two are photometric and spectroscopic surveys of galaxy clusters sampling massive halos, while the latter contains galaxy groups, binary systems, and isolated galaxies. Of the total sample, $\sim$ 66 \% (76/114) of the galaxies belong to galaxy clusters and the remaining 34\% (38/114) to less dense environments, which are much less likely to be affected by ram pressure (RP, hereafter)\footnote{While this is in general true, there are a number of GASP galaxies located in groups that display clear signatures of ram pressure effect.}. For the sake of brevity, we will refer to galaxies non affected by RP as Non-RPS (Non-Ram Pressure Stripped), and to galaxies displaying signatures of RP interaction as RPS (Ram Pressure Stripped).

\section{Bar identification}\label{sec:bar}

Bar detection and characterization was performed by using two different methods. We firstly had visually inspected the images of the 114 galaxies selected from GASP project. As visual detection and classification can in general be subjective and is likely biased towards the presence of strong structures, we also resorted to a more quantitative analysis. To confirm the presence of a bar and measure its length, we implemented the position angle and ellipticity analysis. We present the results of this process in the following subsection.

\subsection{Visual identification}
In order to identify the presence of a stellar bar by visual inspection, we used color images extracted from the spectral datacubes. Exploiting the observed spectral range, the data cubes were collapsed into three different bands, $r$, $i$ and $z$ of the SLOAN photometric system, which were used to create the color images. Then, an inspection was performed independently by three members of the collaboration (Osbaldo S\'anchez Garc\'ia, Bernardo Cervantes Sodi and Giovanni Fasano), with the results been recorded and later compared against each other to set the final classification. Each galaxy was classified as barred or unbarred, identifying ambiguous cases. With this method we found that  20\% (23/114) of the galaxies are barred, $\sim$37\% (42/114) unbarred, and for $\sim$ 43\% (49/114) it was not possible to obtain a clear detection (ambiguous cases), including objects which are observed at a high inclination angle, a feature which notoriously hampers stellar bar identification. An example of each of the types is presented in Figure \ref{galaxies}.

\subsection{Position angle and ellipticity analysis}
\begin{figure}[h!]
\label{distributions}
\centering
\begin{tabular}{c}
\includegraphics[width=0.5\textwidth]{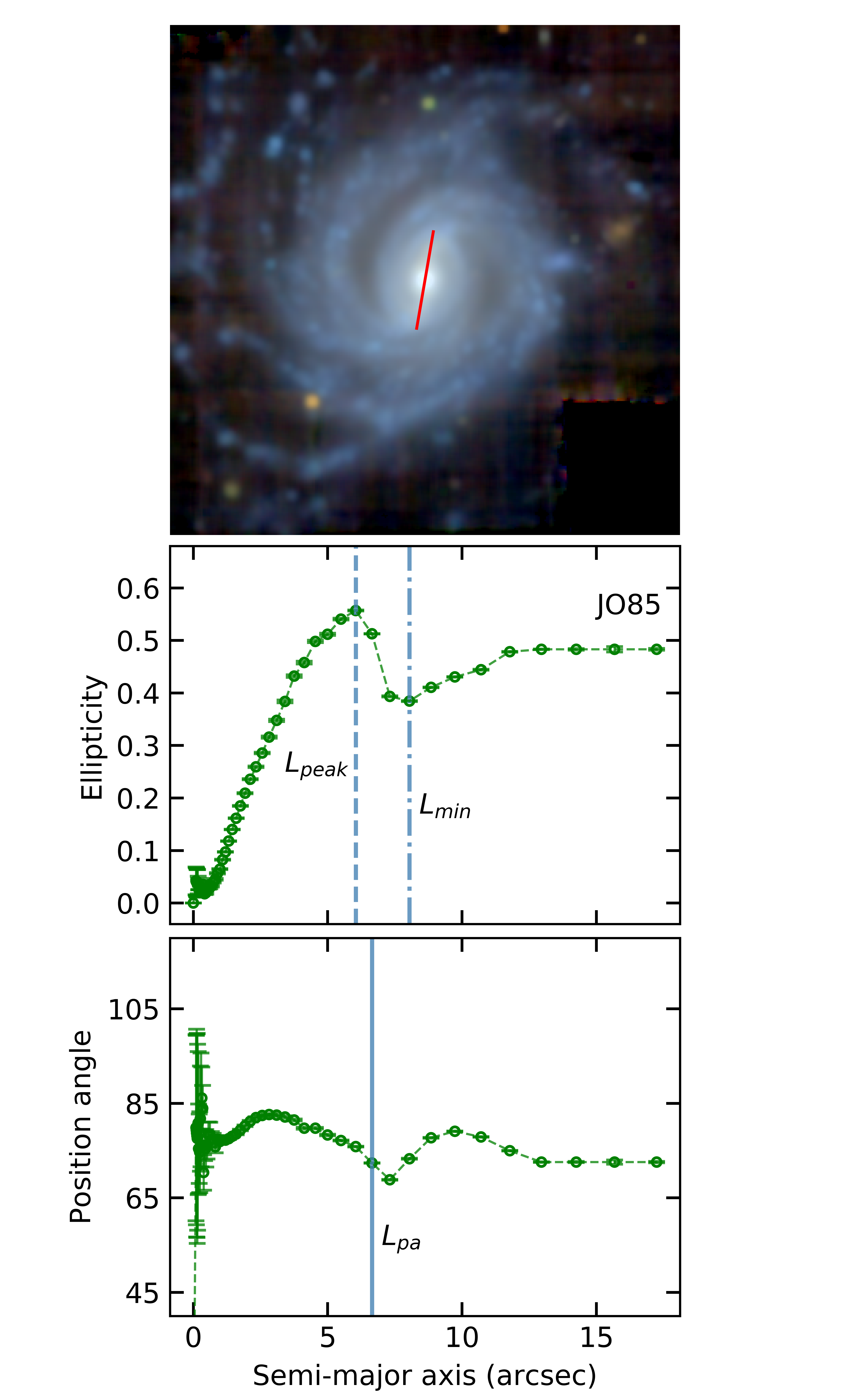} 
\end{tabular}
\caption{\textit{Top panel:} Color image of barred galaxy JO85 with the red line indicating the extent of the bar as measured with Eq. \ref{eq:length}. \textit{Middle panel:} Ellipticity profile of the same galaxy. Dashed line indicates the position of $L_{peak}$ and dashed-doted line indicates the position of $L_{min}$. \textit{Bottom panel:} position angle profile of galaxy JO85, where the solid vertical line indicates the position of $L_{pa}$.
}\label{length}
\end{figure}

In order to confirm the presence of a bar and additionally measure its length, we explored the behavior of the position angle and ellipticity profiles \citep{Wozniak95}. We used \texttt{photutils}, a package from Astropy, to fit elliptical isophotes to the galaxies in the sample. \texttt{photutils} requires an ellipse as an initial value to perform the fit, which is specified by its center, ellipticity ($e$), major axis ($a$) and position angle ($PA$). The center was chosen as the brightest spaxel and the ellipticity ($e$), the major axis ($a$) and the position angle ($PA$) were taken from the values of an ellipse, manually chosen for each object, encompassing the central region of the galaxy. As a result of the adjustment, \texttt{photutils} gives a list of isophotes whose attributes are the fit values for each isophote sorted by the semi-major axis length. With this information we define the position angle, eccentricity and b/a from each galaxy by averaging the values of the three outermost isophotes. A typical $e$ and $P.A.$ profile of a barred galaxy is presented in Figure \ref{length}, where the ellipticity monotonically increases within the bar region, reaching a maximum, while the $P.A.$ remains roughly constant. How abrupt are the changes in $e$ and how constant remains the $P.A.$, depends on how the spiral arms are connected to the edges of the bar, on the presence of a bulge or pseudobulge in the center, and on the orientation of the bar relative to the projected disk.

To measure the length of the bar, one can simply adopt the value of the radius at which $e$ reaches its maximum, although some studies suggest that this value underestimates the true extent of the bar \citep{Wozniak95, Athana02}. Alternatively, other estimates are adopted, such as the first minimum after the maximum in eccentricity or the radius at which a sudden change in $P.A.$ is evident. In this work we adopt a measurement proposed by \cite{Neumann19}, that relies on three different length estimates, namely $L_{peak}$, $L_{min}$, and $L_{pa}$. $L_{peak}$ is defined as the length where the first peak of ellipticity occurs; $L_{min}$ corresponds to the radius where the first minimum in ellipticity is found after the first peak; $L_{pa}$ is defined at a position corresponding to a sudden change in the $P.A.$ after the peak of the ellipticity. For this work, we have taken a value of $\Delta P.A. \sim5^{\circ}$ for this change in $P.A.$, and we denote it as $L_{pa}$. The final estimate for the length of the bar is hence given by:
\begin{equation}
\label{eq:length}
    L_{bar} = AVG(L_{peak},MIN(L_{min}, L_{pa})).
\end{equation}

These criteria allowed us to eliminate the ambiguity found by the visual inspection method confirming  the presence of bars in 35\% (40/114) of the galaxies, while in about the 42\% (48/114) no bar was detected. Figure \ref{length} includes the values of $L_{peak}$, $L_{min}$ and $L_{pa}$ of JO85 as an example, as well as an indication of $L_{bar}$ superimposed to the color image of the galaxy.

\subsection{Sample selection}\label{ssec:sample}

To investigate the combined effect of stellar bars and the ram-pressure phenomena in the resolved star formation history of the galaxies, we start by defining a sub-sample from the original one, in which we are confident that the identification and characterization of the bars is reliable. As already described, the bar identification implemented was through visual inspection of the images and the position angle and ellipticity analysis. For both methods, there is a strong selection bias due to the inclination that strongly favors face-on galaxies. In order to remove this effect, the sample analyzed in this work is restricted to galaxies with minor-to-major axial ratio $(b/a) \geq 0.5$, where the bias is minimized \citep[e.g.][]{Lee+12, munoz-mateos13}. We have used the analysis described in the previous section to refine the criterion for detecting highly inclined objects (edge-on galaxies), which turned out to be the 39\% (44/114) of the initial sample. This further reduces the number of galaxies used, to 33 and 37 for barred and unbarred, respectively.

In table \ref{tab:class} we report the barred vs unbarred classification for the different methods, as well as the number and percentage of galaxies excluded due to the maximum inclination criterion. Finally, we removed four additional galaxies, P96949, JO190, P877, P3984 (two barred and two unbarred). The first one is a merging system between an old gas poor galaxy and an early-type-like one, younger and gas-rich \citep{Vulcani17} and the last tree galaxies are identified by \cite{Vulcani21} as merging systems as well, and including them might introduce some biases in the properties we study. Therefore, our final sub-sample consist of 66 galaxies (35 affected  and 31 unaffected by RPS), mostly face-on, with similar mass distributions within $8.9 \leq$ log$(M_*/M_{\odot}) \leq 11.2$, all presenting late-type morphologies \citep{Poggianti17}. RPS galaxies are members of galaxy clusters, except P96244, and non-RPS galaxies belong to both galaxy clusters and less dense environments. We also note that RPS and Non-RPS galaxies share similar stellar mass distribution.

For the final sample, a fraction of barred galaxies of 47\% is found which, despite the low number of objects here analyzed, is in good agreement with reports from previous studies on samples at low redshift \citep{Barazza08, Aguerri09, Yoshino15}, as well as for the case of galaxies located in clusters \citep{Yoon20}.

\begin{deluxetable*}{cccccc}
\tablenum{1}
\tablecaption{Galaxy Classification by Different  Methodologies}\label{tab:class}
\tabletypesize{\footnotesize}
\tablewidth{0pt}
\tablehead{
\colhead{Classification method} & Number & \colhead{Barred} & \colhead{Unbarred} & \colhead{Highly inclined} & \colhead{Ambiguous}
}
\startdata
Visual inspection & 114 & 23(20\%)  & 42(37\%)  &     --    & 49(43\%)\\
Ellipse fitting   & 114 & 33(29\%)  & 37(32\%)  &  44(39\%) & --  \\
Final sample      & 66  & 31(47\%)  & 35(53\%)  &     --    & -- \\
\enddata
\tablecomments{The final sample consists of barred and unbarred  galaxies with $b/a>0.5$, with the exclusion of 4 galaxies with features associated to tidal/gravitational interactions.}
\end{deluxetable*}

\section{Galaxy properties}\label{sec:SINOPSIS}

Here we present how the analysis for GASP data was performed, with a particular focus on the stellar populations properties. For a more detailed description of the method, tools and adopted parameters we refer the reader to \cite{Poggianti17}.

\subsection{Stellar population properties derivation}
The data of the 114 galaxies observed with MUSE were reduced following the standard reduction procedures as described in the pipeline manual \footnote{http://www.eso.org/sci/software/pipelines/muse} \citep{Bacon10}. The full treatment of the data cubes has been described in detail in \cite{Poggianti17}, and here we will only briefly recall the adopted steps. The cubes are first corrected for extinction due to our own Galaxy, using the extinction value estimated at the galaxy position \citep{Schlafly11} and assuming the extinction law from \cite{Cardelli89}. Hence, stellar and ionized gas redshift are derived on a spaxel-by-spaxel basis by means of the Penalized Pixel-Fitting (pPXF) code \citep{Cappellari04} and of the  KUBEVIZ software \citep{Fossati16}, respectively. These are used to derive the stellar and gas kinematics, and hence feed to the spectrophotometric fitting code {\sc sinopsis} \citep{fritz07,fritz17} to derive stellar population properties on a spatially resolved base. The code also produces pure stellar emission models which are hence used to subtract the underlying stellar absorptions and create an ``emission line-only'' cube. Then, KUBEVIZ is run once again on the latter datacube, to calculate line fluxes, that are hence used to create the final line emission datacube, in which internal dust extinction is corrected, by means of the Balmer decrement, and assuming again the \cite{Cardelli89} extinction law. 

Given that we extensively use the output datacube of {\sc sinopsis}, such as the stellar mass and ages for our analysis, here we give a very brief summary of its most important features. We refer the reader to \cite{fritz17} for a more complete description of the characteristics, assumptions, and working mode of {\sc sinopsis} applied to GASP spectra. 

As already mentioned, {\sc Sinopsis} is a spectrophotometric code which, by means of simple stellar population models (SSP) reproduces the spectra in each spaxel of a MUSE datacube within the observed galaxy. In all of the GASP works, we have used SSP spectra from the newest model grid of \textcolor{blue}{Charlot \& Bruzual (in prep.)}, assuming a \cite{chabrier03} initial mass function (IMF). 12 different age values have been adopted, obtained by re-binning models from an initial grid of 220 ages, going from $10^4$ to $14\times 10^9$ yr. When relevant, nebular emission lines were also calculated by ingesting the spectra into the photoionization code {\sc cloudy} \citep{ferland98,ferland13}. These affect the spectra up to ages of about $\sim 2\times 10^7$ yr, and allow to reproduce not only pure stellar features, but also emission lines (namely H$\alpha$ and H$\beta$ for GASP spectra). 

Due to well known degeneracy issues, we cannot rely on the star formation history (SFH) in the 12 age bins. This is why the age resolution has been further reduced, by further re-binning the 12 ages into 4, providing more robust and reliable results. These were chosen in such a way that the spectral differences between the stellar population within each of them were maximized. They are defined in these age range: $0 - 2\times 10^7$, $2\times 10^7 - 5.7\times 10^8$, $5.7\times 10^8 - 5.7\times 10^9$, $5.7\times 10^9 - 14\times 10^9$.

To summarize, the mass and age of the galaxies were obtained with {\sc sinopsis} and the flux and equivalent width of the emission lines with KUBEVIZ. 

Star formation rate maps were calculated from H$\alpha$ luminosities on a pixel-by-pixel basis, correcting for both underlying photospheric absorption and internal dust extinction \citep[see e.g.][for further details]{Poggianti17}.

\subsection{Ram-pressure stripping stages}

Galaxies affected by ram pressure stripping are most likely objects that are in their first phase of falling onto a cluster and, according to the \cite{GunnGott72} parametrization, the pressure force felt by the interstellar medium of such galaxies depends on the density of the intracluster medium and the velocity at which the galaxy falls. 
Depending on its mass, morphology, inclination between the disk of the galaxy and its velocity, ram pressure might affect infalling galaxies differently. Furthermore, the jellyfish morphology takes a certain amount of time to develop. Numerical hydrodynamical simulations, for example, have shown how the gas is firstly removed from the galaxies' outskirts, and stripping hence moves inward \citep{Steinhauser12}. Once a galaxy is stripped of most of its gas and it is past its peak stripping phase, ionized gas tails will fade out, and what is left is a galaxy with a truncated disk or, more in general, with a much lower gas content. 

This can be used to broadly and roughly define the stage of interaction that galaxies which are (or have been) subjected to ram pressure find themselves into, and relate it to the spatially resolved properties of the stellar populations in particular.

This was assessed by a visual inspection of H$\alpha$ maps and, as a result, galaxies have been divided into 4 categories: initial, stripping, jellyfish and truncated, where the effect of the interaction grows gradually from the first to the last case.

\subsection{AGN identification}
 
\begin{figure}[!t]
\label{decomposition}
\centering
\begin{tabular}{c}
\includegraphics[width=0.5\textwidth]{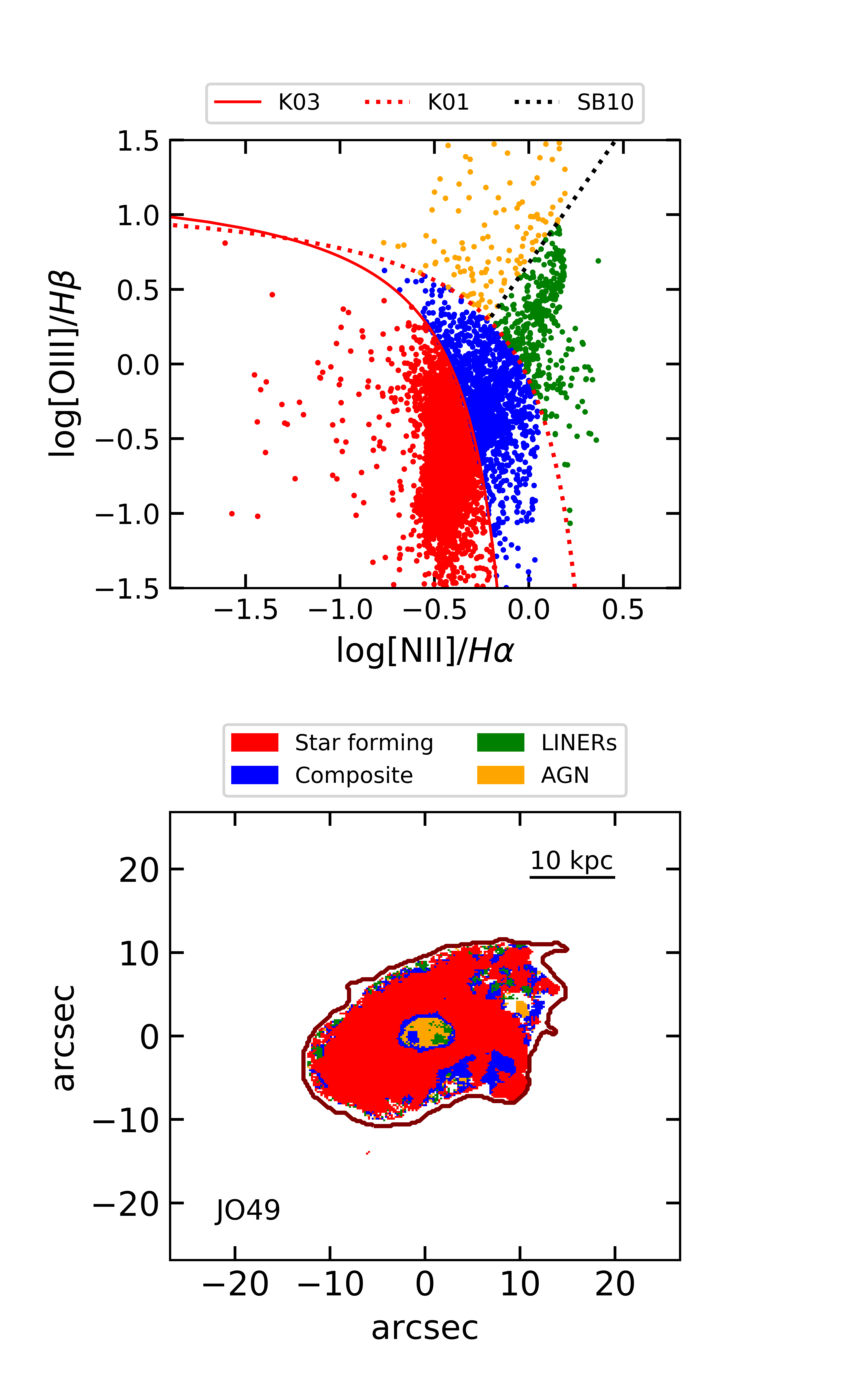} 
\end{tabular}
\caption{\textit{Top panel}: BPT diagnostic diagrams for JO49 galaxy using {\sc [Oiii]}5007/$H\beta$ versus {\sc [Nii]}6583/$H\alpha$ line ratios for all
spaxels with fluxes measured with an S/N $>$3. \textit{Bottom panel}: corresponding map showing the ionization mechanism as derived from the BTP diagram. The maroon line delimits the most external stellar isophote.}
\label{bpt}
\end{figure}

In order to take into account the fact that gas ionization processes can come from mechanisms other than photoionization from high mass stars, we have used \citep[][BPT]{Baldwin81} diagrams based on the {\sc [Oiii]}5007/$H\beta$ versus {\sc [Nii]}6583/$H\alpha$ lines ratios. The classification of the star forming regions, composite, LINER/AGN was adopted from the results of \cite{Kewley01} and \cite{Kauffmann03}. A diagnostic diagram of JO49 is shown as an example as well as its corresponding spatially resolved classification in Figure \ref{bpt}. As an additional constraint, we discard those spaxels with EW($H\alpha) \leq 3$ (in emission), in order to remove possible contributions from other ionization sources such as weak AGNs or shocks \citep{Cid10, Cid11}. 

Relying on these diagnostics, ionized gas emission was accordingly classified on a spaxel--by--spaxel base and, as far as SFR maps are concerned, only spaxels classified as ``purely star forming'' were used. Exploring the occurrence of AGN activity in ram-pressure stripped galaxies in local clusters, \cite{Peluso22} identified twelve galaxies hosting an AGN from GASP sample. Six of them had already been reported by P17, one galaxy was studied by \cite{fritz17}, and other five turned out to be previously undetected AGNs. We use the AGN identification from that work since their detection strategy is more reliable, having been complemented by the use of additional information, such as X-ray emission, to confirm the presence of AGN.

As already stated previously, to identify the presence of a stellar bar we require that the galaxy be face-on. So, of the 12 GASP AGN hosts, only six are face-on (JO49, JO85, JO171, JO194, JO201, JW39). All of them host a stellar bar, with the  exception of JW39\footnote{Note that this galaxy shows hints for the presence of a stellar ring} and JO171. Furthermore all of them are strongly affected by ram pressure.

\section{Results and discussion}\label{sec:Results}

\subsection{Radial profiles}\label{sec:profiles}

To study the combined effect of ram-pressure and the presence of bars in the star formation history of our galaxies, we produced spatially resolved maps  of different indicators of the present and past star formation activity, which we describe and motivate in the following. The star formation rate surface density, $\mathrm{\Sigma_{SFR}}$, calculated from the $\mathrm{H\alpha}$ luminosity as explained above), was used as this allows a comparison with previous works \citep[e.g.][]{Chown19, Lin20}. For the same reason, the equivalent width of the $\mathrm{H\alpha}$ line --$\mathrm{EW(H\alpha)}$-- was also adopted plus, this has the advantage of being a good proxy for the specific SFR (sSFR; star formation rate per unit stellar mass), which allows a more unbiased evaluation of the recent star formation activity (i.e. independent on the stellar mass). As both stellar mass and SFR are available from our dataset, we also used sSFR maps, as a more direct measure when compared to $\mathrm{EW(H\alpha)}$. Finally, we used the luminosity weighted age, directly calculated from the spectral fitting with SINOPSIS.

From each one of the aforementioned maps, we have then estimated the one-dimensional radial profiles of each parameter correcting for effects of inclination, by taking into account the global ellipticity and position angle. The profiles are created using concentric elliptical annuli with a spatial sampling of 1 arcsec along the semi-major axis and whose common center is the brightest spaxel at the center for each galaxy. The position angle and ellipticity of the elliptical annuli are kept constant using  global values. Using spaxels with signal-to-noise greater than 3, we calculate the average value within each annulus of $\mathrm{\log\Sigma_{SFR}}$, $\mathrm{\log (EW(H\alpha))}$, $\mathrm{\log (Age)}$. For those spaxels that are intersected by the rings, only the fraction that falls within the annulus in question is taken into account. Examples of the two-dimensional maps and one-dimensional radial profiles for two galaxies of the sample are shown in Figure \ref{radial}.

The profiles extend out to the region contained within the mask that delimits the stellar disk of the galaxy (therefore, we excludes the gas tails from our analysis in those galaxies where they are present), as defined by \cite{Gullieuszik20}. Figure \ref{radial} presents two cases of galaxies with similar stellar mass and two different stages of the ram-pressure phenomena, with the additional difference of JO73 being a barred galaxy, while JO180 is classified as unbarred. In the top panels we show the color image of the corresponding galaxy and maps for $\mathrm{\log\Sigma_{SFR}}$, $\mathrm{\log EW(H\alpha)}$ and $\mathrm{Age}$ ( we do not show any example for the sSFR, as they very closely mimic the ones of $\mathrm{H\alpha}$). In the bottom panels we present the mask that encompasses the stellar disk of the galaxy and the radial profiles of the same parameters. Finally, 8 galaxies do not display any star formation activity, and hence lack emission lines. For these ones, no SFR or EW(H$\alpha$) profile could be created and, in these cases, we relied on the average stellar age only. Taking this into account, we are left with a total of 62 galaxies with all three radial profiles. In the case of the luminosity weighted age profiles, we count on this information for all the galaxies in our sample, since the age estimation is performed by means of SINOPSIS.

\subsection{Identification of the “turnover” feature}

To measure the average radial change in the star formation (the so--called \textit{``turnover'' feature}) within the central region of a galaxy (i.e. where a bar is likely to have an influence), we carry out a similar procedure to the one described in \cite{Chown19}, \cite{Lin17} and \cite{Lin20}. We inspect the inner region of the profiles of each galaxy and judge whether or not the central region shows a change in the trend of the profile. If such change is identified, we define the turnover radius ($r_t$) as the galactocentric distance at which this occurs.

\begin{figure*}
  \begin{center}
    \includegraphics [width=0.9\hsize]{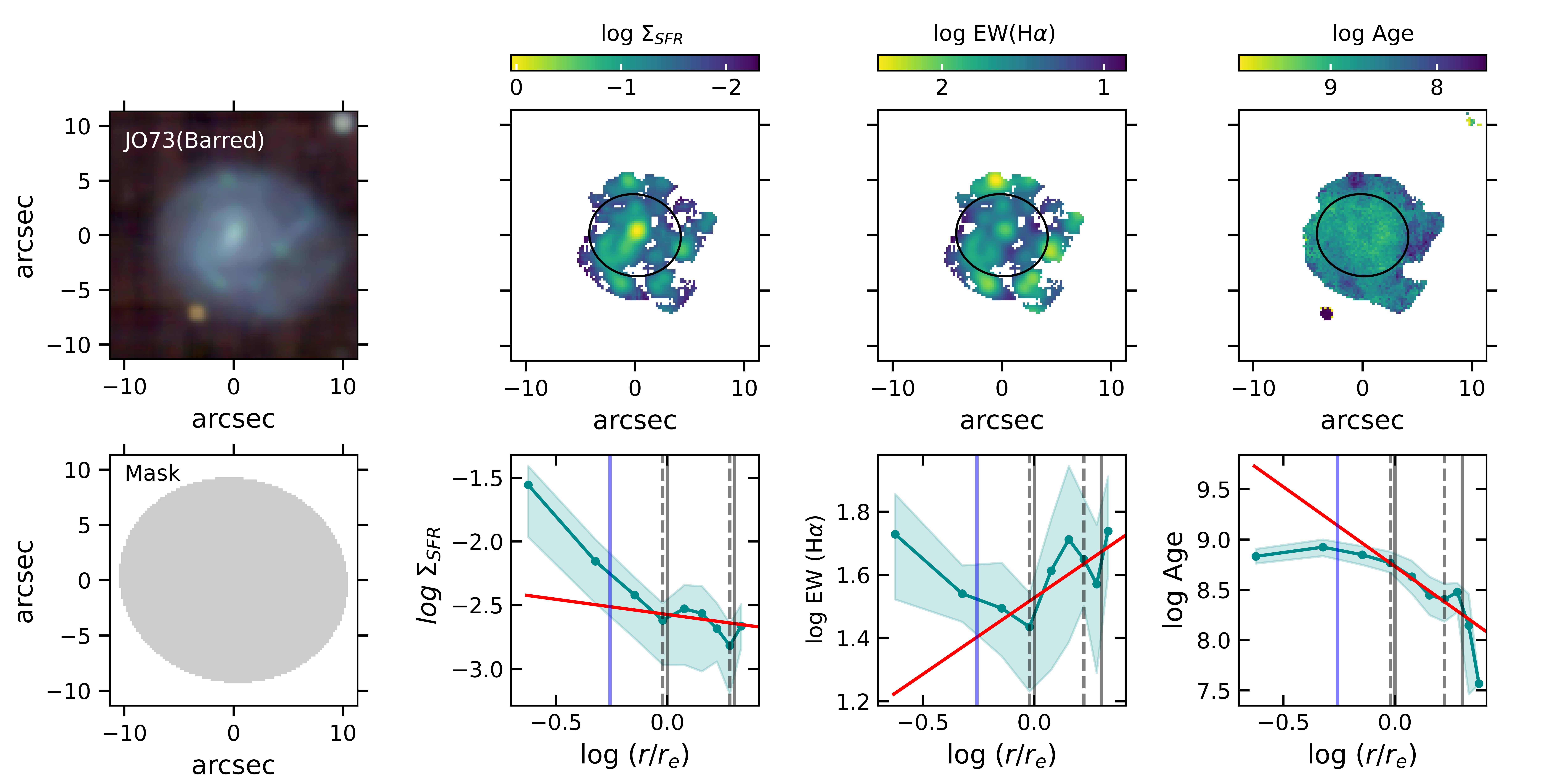}   
    \includegraphics [width=0.9\hsize]{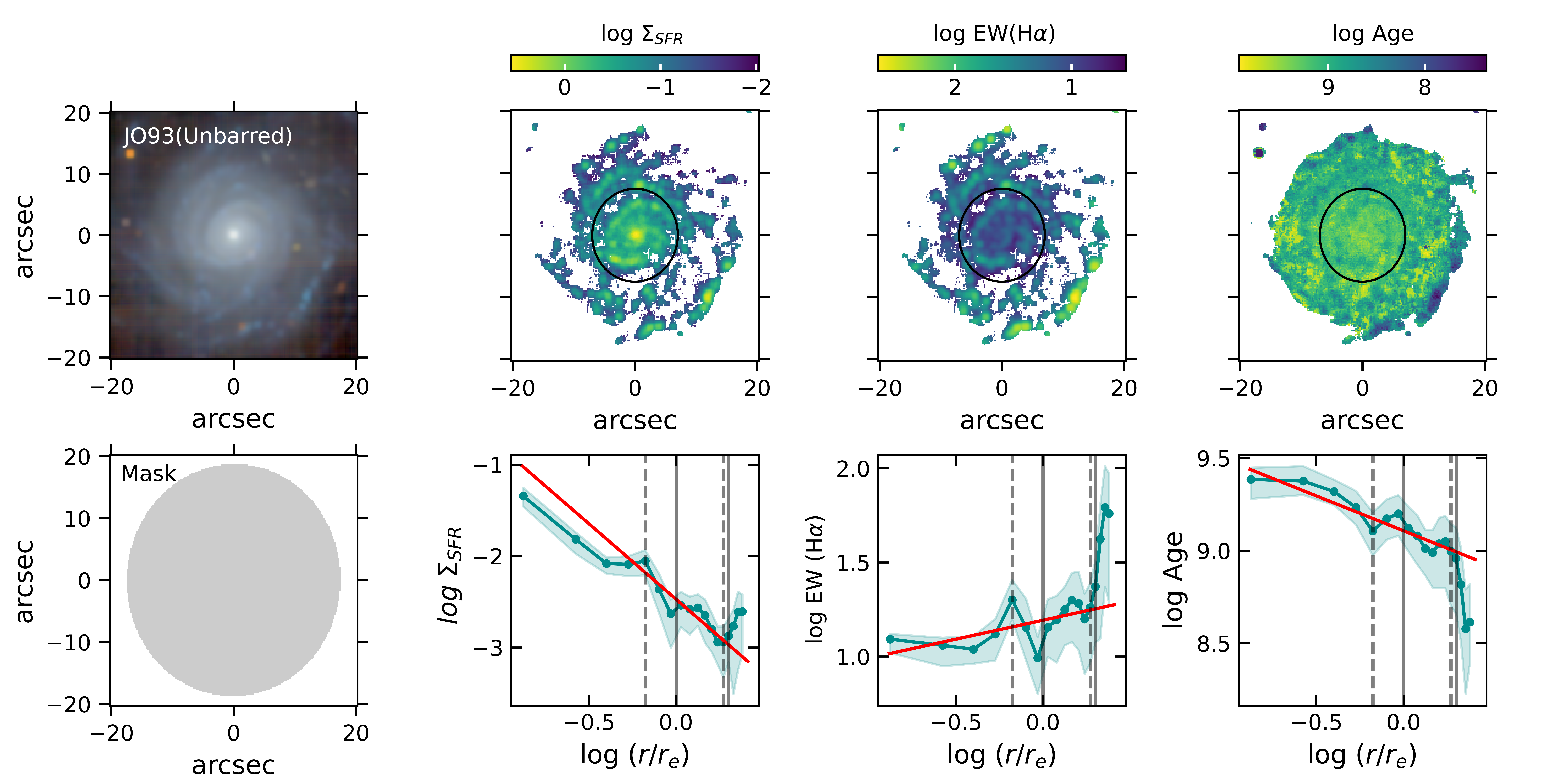}   
    \caption{Maps and radial profiles of the JO73 barred galaxy (\textit{top panel}) and JO93 unbarred galaxy in our sample. For each galaxy, the upper panels show the  optical image followed by the two-dimensional maps of $\mathrm{\log\Sigma_{SFR}}$, $\mathrm{\log EW(H\alpha)}$ and $\mathrm{Age}$. The color coding shown on the maps represents the different values taken by the three indicators of recent star formation activity within the galaxy, as indicated by the colored bars at the top of the plot. The lower panels (from left to right) show the mask of the galaxy followed by the one-dimensional radial profiles of $\mathrm{\log\Sigma_{SFR}}$, $\mathrm{\log EW(H\alpha)}$ and $\mathrm{Age}$. The vertical dashed lines indicate the region where the linear adjustment was made in the radial profiles. The vertical gray lines represent one and two effective radius (the one $r_e$ coincide with the corresponding ellipse in the two-dimensional maps). The vertical blue line represents the length of the bar.}
  \label{radial}
  \end{center}
\end{figure*}

Since at least half of the galaxies in the sample are affected by RP, we calculate the effective radius ($\mathrm{ r_e }$) in order to make sure that the  turnover radius (which is identified by visual inspection) is within the effective radius. The fact that the radius of the turnover is within the effective radius of the galaxy helps to rule out changes in the slopes of the profiles that are caused by the ram-pressure affecting the outer parts of the galaxies that present this phenomenon. Figure \ref{radial} shows the effective radius with a gray vertical solid line for each galaxy profile.

\begin{figure*}[!t]
  \begin{center}
    \includegraphics [width=0.9\hsize]{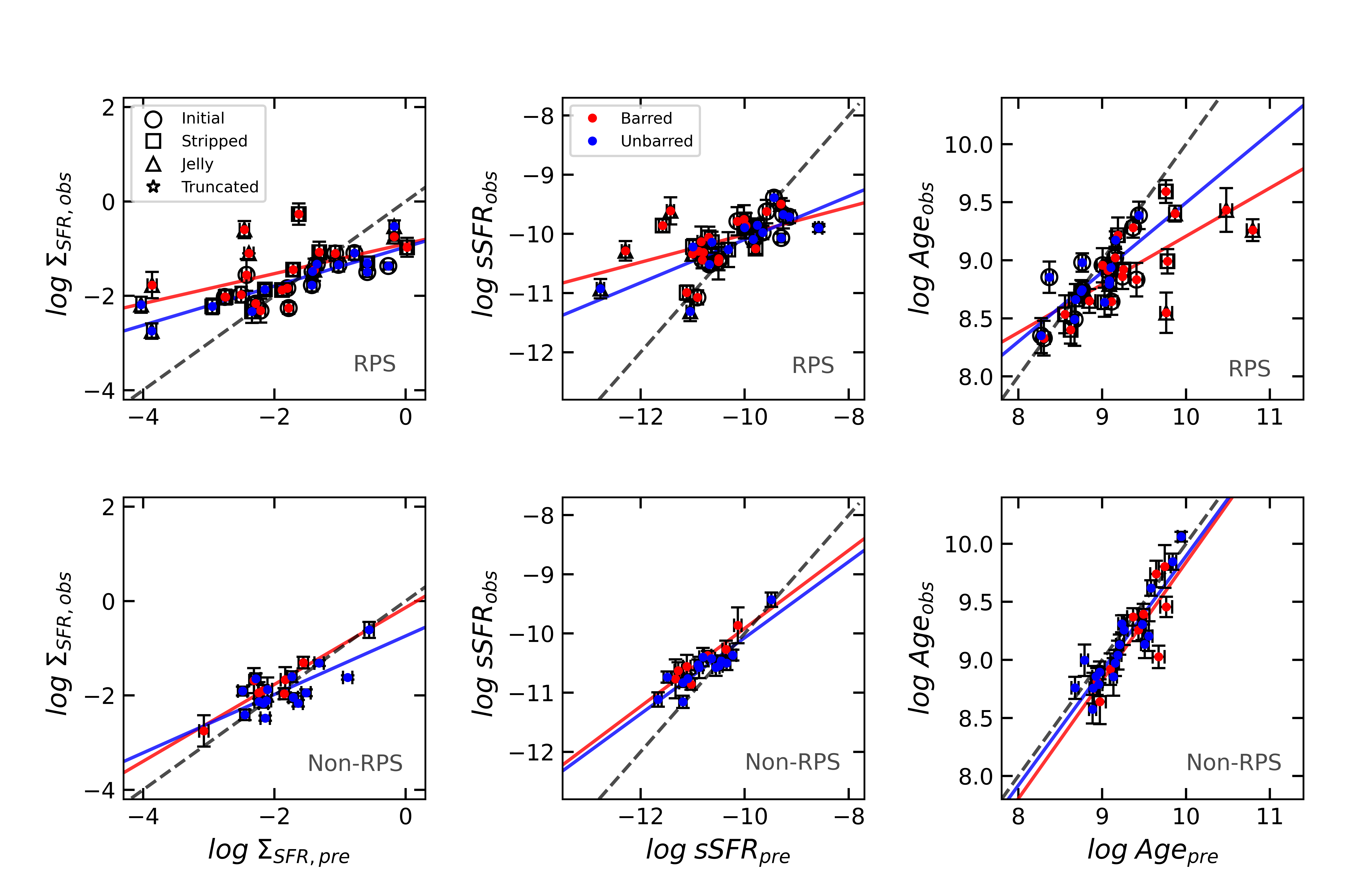}   
      \caption{Comparison between the observed central and extrapolated values of the three SFH indicators with their respective error bars. The 1:1 relation is shown as a dashed black line,  Red points represent bar galaxies and blue points unbarred ones. The solid red and blue  lines correspond to fits for the respective sub-samples. The different unfilled markers in the top panel represents four different stages of ram pressure; initial, stripping, jellyfish and truncated, where the effect of the interaction grows gradually from the first to the last case.} 
  \label{comparasion}
  \end{center}
\end{figure*}

Finally, the identification of the turnover radius for each SFH indicator is  performed by visual inspection, however, not independently. Once the change in slope has been identified in one SFH indicator (e.g., $\mathrm{\Sigma_{SFR}}$ ), we compare the radius at which this occurs with the radius identified in another SFH indicator (e.g., $\mathrm{EW(H\alpha)}$, sSFR, or/and $\mathrm{Age}$), if  the values found in this way are similar we proceed to store the  corresponding value. When performing the analysis to present our results, we found that the trends resulting from the sSFR and the $\mathrm{EW(H\alpha)}$ maps, are basically undistinguishable. Of course, this is expected as $\mathrm{EW(H\alpha)}$ is a proxy for the sSFR. Furthermore, the results for sSFR are very similar to those obtained with $\mathrm{\Sigma_{SFR}}$ (see also Fig. \ref{comparasion}). For this reason, in the following sections we will mostly focus on results solely based on the $\mathrm{\Sigma_{SFR}}$ and $\mathrm{Age}$ indicators, which also allows us for a more fair comparison with results in the literature, that have used the same SFR indicators.

\subsection{Central Star Formation}\label{sec:Central_Star_Formation}

To measure the turnover strength, we carry out a procedure similar to the one described in \citet{Lin20}. For each galaxy, and for each given SFH indicator, we start by fitting a line to the radial profile between $\mathrm{r = r_t}$ and $\mathrm{r = r_t + 3}$ arcsec. We then check whether the next data point follows the same trend, evaluating whether the line of best fit passes within the error bar of the point. If so, we perform the linear fit again, but now adding the new point  (that is, $\mathrm{ r_t + 3}$) and so on until we find that this condition is not met anymore, or until we reach two effective radii. We chose to make the fit within this range to analyze the trend of the radial profile without it undergoing a change in the slope in the central region. In Fig. \ref{radial}, the radial ranges of the linear fits are plotted as gray dashed lines for two example galaxies. 

For each of the SFH indicators we quantified its turnover strength as the difference between the observed and extrapolated values in the central region, as measured by \citet{Chown19}. The turnover strength is defined as:
\begin{equation}
\Delta Y \equiv Y(r=0) - Y_{extrap}(r=0),     
\end{equation}
where $\mathrm{Y(r = 0)}$ is the value of $\mathrm{Y}$ in the central radial bin, and $\mathrm{Y_{extrap}(r = 0)}$ is the best-fitting line extrapolated to $r = 0$. In order to investigate the combined influence of the ram-pressure phenomena and the presence of a bar on the turnover nature of our galaxies, we segregated our sample into those with and without evidences of ram-pressure stripping (RPS) effects, and between barred and unbarred galaxies. Then, we proceed with the comparison between the observed central value and the extrapolated ones of the two SFH indicators. The result is displayed in Figure \ref{comparasion}. Top panels show RPS galaxies and bottom panels Non-RPS galaxies with red points representing barred galaxies while blue indicating unbarred ones. For all panels, the dashed line represents the 1:1 relation.The x-axis error bars denote the estimated 1-sigma confidence interval around the adjusted extrapolated profile, while the y-axis error bars indicate the dispersion around the mean value for the inner most region of the corresponding parameter. For RPS galaxies, the different stripping stages are shown using different markers. Finally, the red and blue solid lines correspond to separately fitting the sub-samples of the barred and unbarred galaxies respectively. We point out that in Figure \ref{comparasion} we only include 58 galaxies with estimated $\mathrm{\Sigma_{SFR}}$ and sSFR, while for the plot regarding $\mathrm{Age}$ we count with 66 galaxies due to the reasons described in section \ref{sec:profiles}.
\begin{figure*}
  \begin{center}
    \includegraphics [width=0.9\hsize]{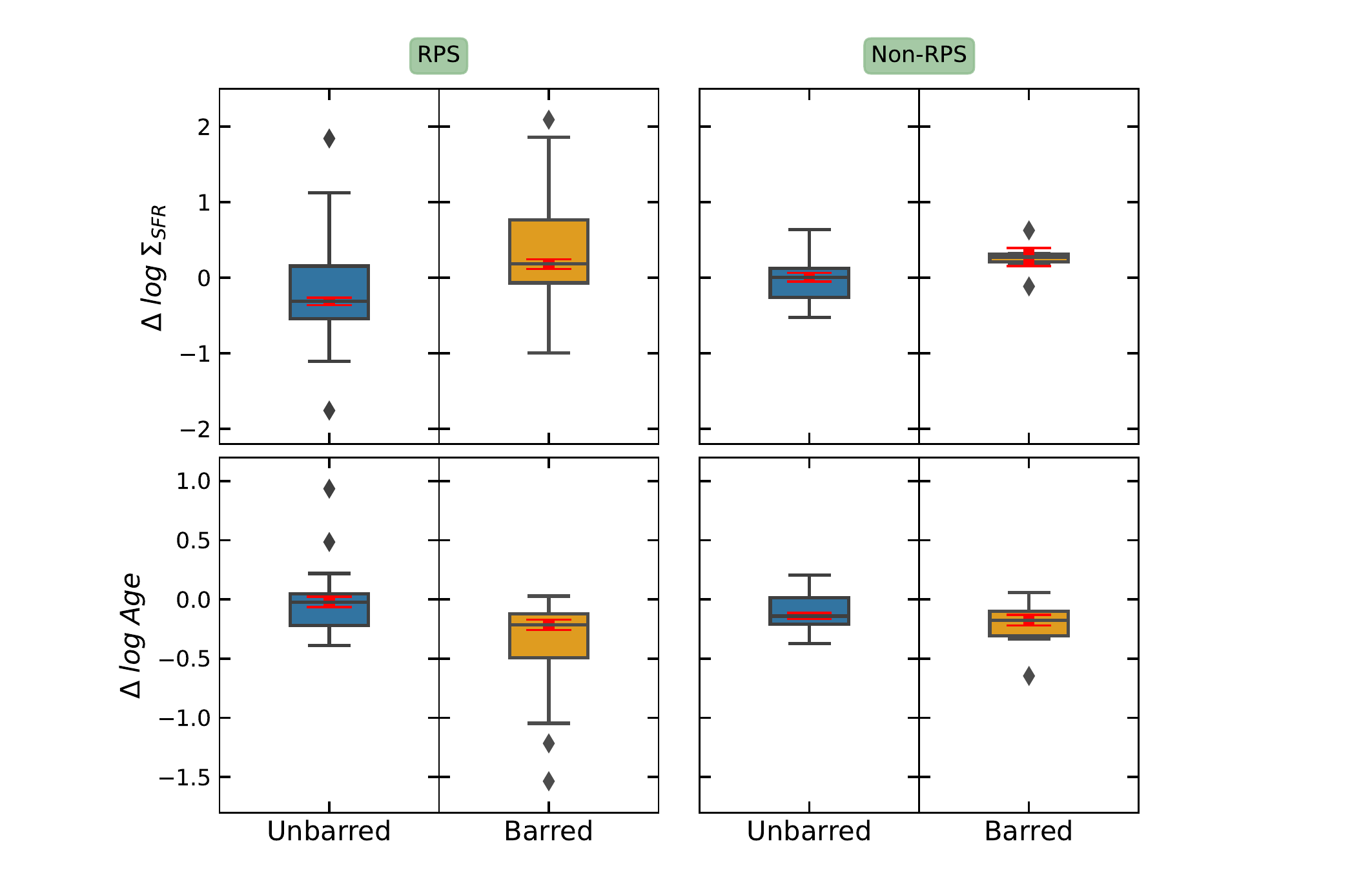}   
      \caption{Turnover strength for $\mathrm{\Delta \log\Sigma_{SFR}} $ and $\Delta  Age$ of RPS (left panels) and Non-RPS (right panels) galaxies, segregated into barred and unbarred galaxies. Boxes mark the interquartile range of each sample, the horizontal black line in the boxes represents the median value. The top and bottom black line stand for the highest and lowest value, respectively. The dotted diamonds are the outliers for each sub-sample. Here we also report, in red, 4$\sigma$ values calculated by bootstrap.}
  \label{box}
  \end{center}
\end{figure*}

From Figure \ref{comparasion} we observe that the RPS galaxies (top left panel) have a larger scatter (around  of the 1:1 relation) than the Non-RPS ones (bottom left panel) by 0.1dex. This is also shown in the top panels of Figure \ref{box} where we present box plots for the turnover strength for the star formation rate surface density: the distributions of galaxies experiencing ram pressure are more extensive with respect to their counterparts.  As the sizes of the sub-samples are small, we implemented a bootstrap resampling method to test the statistical significance of our results. The test relied on the generation of 1,000 random realizations from the original sample and the variance of the central value of each accounted variable. As an outcome of this procedure, we also show as red bars the 4-sigma confidence intervals, based on the random realizations of the turnover strength values. 

We also note that more than half of both RPS and Non-RPS galaxies (left panels) are above the 1:1 relation. This is also shown in the top panels of Figure \ref{box}, where the median of the distribution falls on positive values, indicating that the average population of galaxies experiences an enhancement in the star formation rate in the central regions. Furthermore, as expected based on the results of \cite{Vulcani20}, we note that this enhancement is slightly higher in RPS galaxies. 

When we segregate our sample into barred and unbarred galaxies, we note that this enhancement is larger for barred galaxies than for unbarred ones. Nevertheless, this difference becomes even more noticeable if these galaxies are undergoing ram-pressure stripping, by a factor of 0.15dex. This is also reflected in the median values in the top left panel of Figure \ref{box}, where box plots for barred and unbarred galaxies undergoing ram-pressure stripping are shown. Finally, we observe that RPS galaxies that host a bar are those that present the furthest distance from the 1:1 relation (i.e., galaxies that are beyond the 1-sigma distribution), while for Non-RPS galaxies there is no strong distinction. When we take into account the stripping stage of the galaxies that present the greatest distance of the 1:1 relation, we observe that they are galaxies that  show the highest degree of disturbance in the ionized gas distribution, and are classified as ``stripping'' or ``jellyfish'' galaxies.

\begin{figure*}
  \begin{center}
    \includegraphics [width=0.9\hsize]{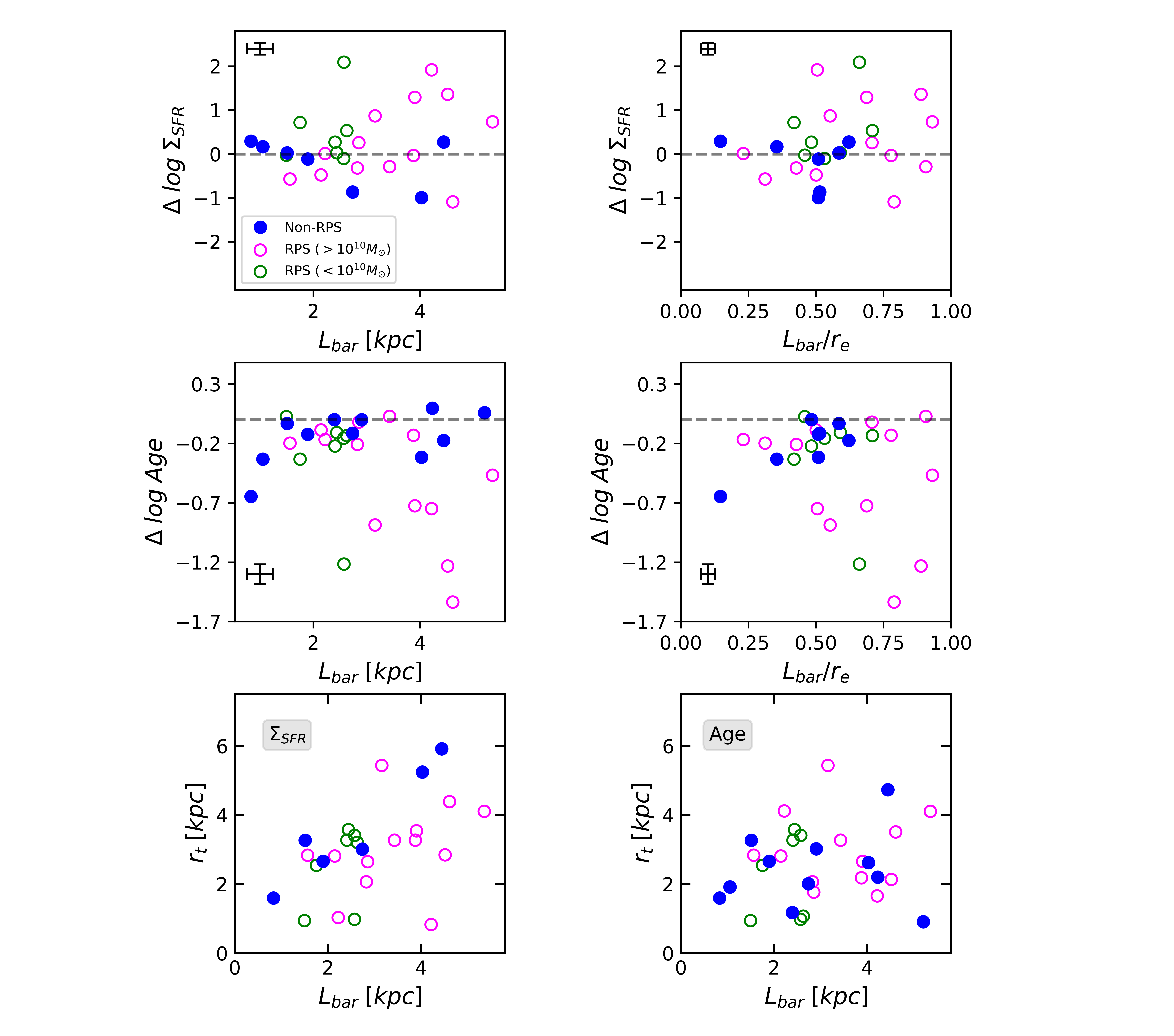}   
      \caption{\textit{Top panels}: $\mathrm{\Delta \log\Sigma_{SFR}} $ turnover strength as a function of bar radius (left) and normalized bar radius (right). \textit{Middle panels}: $\Delta \log Age$ turnover strength as function of bar radius (left) and normalized bar radius (right). Lower panels: turnover radius as function bar radius (right) and normalized bar radius (right).  Non-RPS galaxies  are indicated with blue points and  RPS galaxies  are indicated with open circles (green open circles represents galaxies with log $  (M_*/M_{\odot}) < 10$ and the magenta ones with log$(M_*/M_{\odot}) \geq 10$). Typical error bars for individual galaxies are shown in the top left for top panels and bottom left for middle panels.}
  \label{lengths}
  \end{center}
\end{figure*}

Regarding the behavior of the average age of the stellar populations in the central region, we find a similar trend between RPS and Non-RPS,  with the former being more scattered around the 1:1 relation by $\sim 0.1$ dex (right panels) with respect to the latter. This trend can be better appreciated in the bottom panels of Figure \ref{box}, where the distributions in the box plots of RPS galaxies (left-bottom panels) are more extended than Non-RPS ones (right-bottom panels). Besides, when we identify the galaxies that are above and below the 1:1 relation, we find that more than half of the RPS galaxies are below the relation, while for Non-RPS galaxies we find no difference. Therefore, these results show that the typical galaxy of our sample shows a rejuvenation in the central region when it is undergoing ram-pressure stripping.  These trends are common to all the SFR indicators and to the average stellar age as well, and can be also appreciated in Figure \ref{box}, where the medians of the distributions of the RPS galaxies present positive values, while the medians of the Non-RPS galaxies are practically at 0. When we segregate between barred and unbarred galaxies we do not find a difference between the average age of the stellar populations in the central region of the Non-RPS galaxies, nevertheless, we do find a difference in RPS ones, finding that barred galaxies tend to have a more pronounced rejuvenation and a stronger increase of the star formation activity than unbarred ones. This can be seen in Figure \ref{box} when comparing the medians of the distributions in the bottom panels. Finally,  similarly to what we found with the analysis of the SFR, we observe that RPS galaxies that host a bar are those that present the furthest distance from the 1:1 relation (i.e.,  are those that lie beyond the 1-sigma line of the distribution), while for Non-RPS galaxies there is no strong distinction. When taking into account the stripping stage of the galaxies, we identify that they are  the objects presenting the strongest stripping evidences.

\subsection{Relationship between the turnover feature and bars length}\label{sec:turnover_length}

Another interesting aspect to analyze is the dependence of the turnover strength as a function of the length of the bar. Since bars promote gas funneling towards the center of the galaxy, we expect the bar length to play a role on the stellar populations rejuvenation in the central region. In principle, a longer bar will sweep a larger disk area, being hence able to perturb and convey towards the center a larger amount of gas \citep{Geron21} and produce a stronger increase of the star formation activity, although this correlation between the magnitude of the enhancement and the length of the bar is generally not found in observational studies \citep{Ellison+11}. The upper and middle panels in Figure \ref{lengths} show the turnover strength accounted by the two different indicators, as a function of the length of the bar (left panels) and the length of the bar normalized to the effective radius of the host galaxy (right panels). In the same figure, lower panels, we show the dependence of the turnover radius on the length of the bar. RPS and Non-RPS galaxies are indicated by open and filled circles respectively. Green open circles represent massive galaxies (log$(M_*/M_{\odot}) < 10)$ and magenta open circles the less massive (log$(M_*/M_{\odot}) \geq 10)$ ones.

In the case of Non-RPS galaxies, represented with blue points in Figure \ref{lengths}, when uncertainties are taken into account, we do not find a variation of the turnover strength for any of the SFH indicators, as function of bar length. In the case of RPS galaxies instead (open circles), we do find a appreciable variation as function of the bar length. When we segregate RPS galaxies into low and high mass, we note that low mass galaxies have a shorter bars than high mass galaxies and they systematically present an enhancement in the star formation rate (positive values of $\mathrm{\Delta \log\Sigma_{SFR}} $ and a rejuvenation for the case of ages in the central regions of the galaxies (negative values of $\Delta$ log Age). In the case of high mass galaxies we find both, a combination of enhancement and suppression in the central regions, with a larger scatter than for the case of low mass galaxies.

The fact that all low-mass galaxies have an enhancement in star formation rate in the central region can be attributed to their higher gas fraction than their more massive counterparts, as suggested by the $M_{*}$-$M_{HI}/M_{*}$ relation \citep{Catinella10,Huang12}. Bars in these low-mass galaxies are also younger than in their high-mass counterparts, as they reached the required gravitational mass to sustain the formation and growth of the bar more recently and hence, the effect of the bar growth is still at work with the available gas, stimulating the star formation activity in the center and lowering the average age of the stellar population in this region. The combined effect of ram pressure and the presence of a bar could make this process more efficient, since the gas redistribution by the ram pressure could replenish the required gas to interact with the bar, specially in low-mass galaxies due to their gas reservoirs being richer  than in high-mass ones. In this way, it would be more likely to find a low-mass galaxy with star formation activity than in a high-mass galaxy where they could be more intermittent due to their lower gas content.

\begin{figure*}
  \begin{center}
    \includegraphics [width=0.9\hsize]{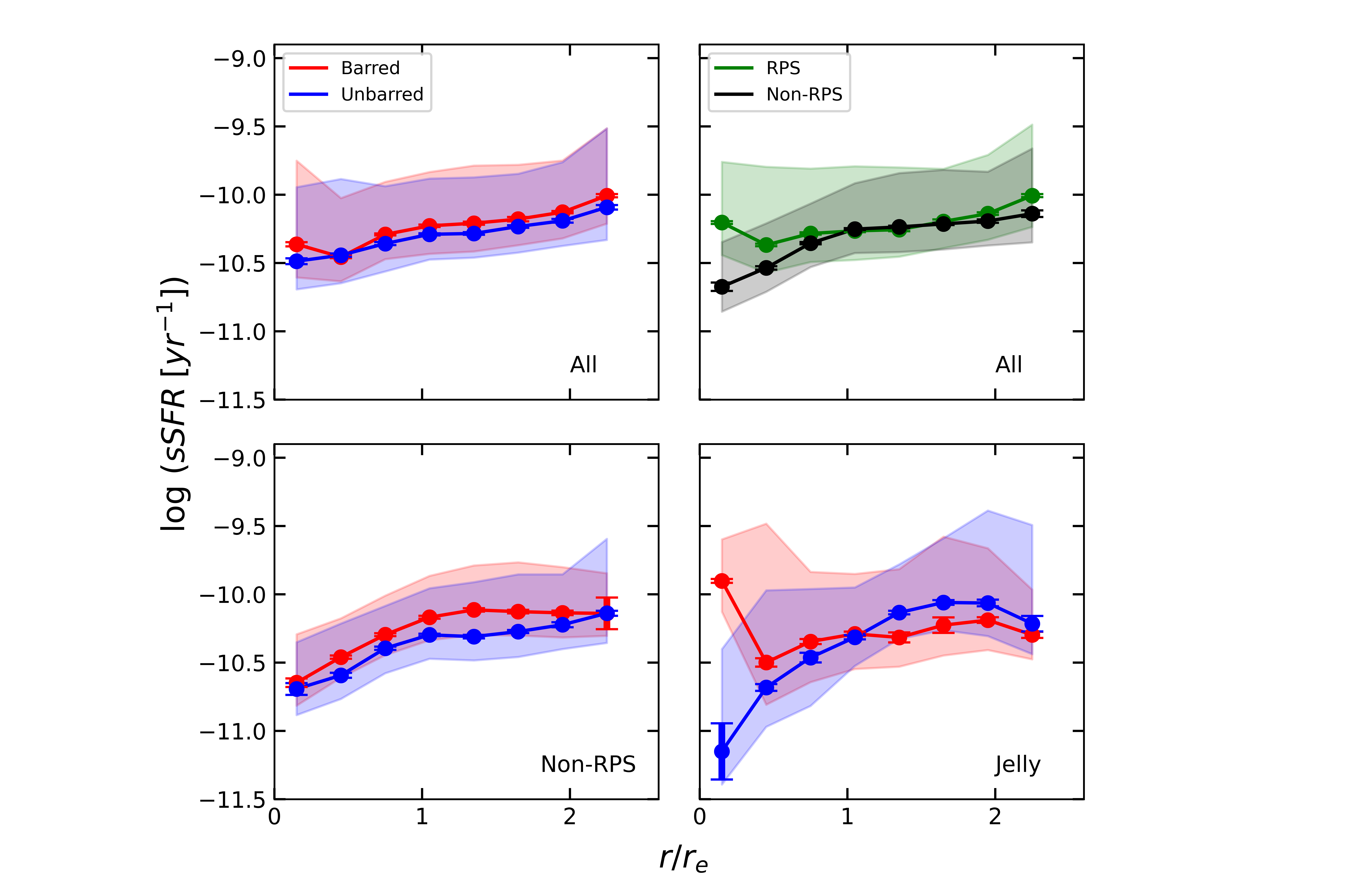}   
    \caption{Specific star formation rate as a function of the normalized galactocentric radius.  Blue and red symbols represent barred and unbarred galaxies, respectively. Green symbols represent RPS galaxies, while black symbols represent Non-RPS galaxies. \textit{Top panels} include all galaxies: \textit{left panel} segregates barred and unbarred, while \textit{right panel} segregates  RPS and Non-RPS. In the \textit{lower-left panel}, we separate galaxies that are not affected by RP by bar presence, and in the \textit{right-panel} we repeat the same for galaxies with RP signatures. Points in each profile represents median values. The shaded regions enclose the interval between the 25th and 75th percentiles, while the error bars denote the dispersion around the median obtained from a bootstrap re-sampling test.}
  \label{specific}
  \end{center}
\end{figure*}

We also explored if our sample presents a correlation between the  turnover radius and the  bar length, as the one reported by \cite{Lin17} and \cite{Lin20}. Bottom panels of figure \ref{lengths}, present the turnover radius as a function of bar length, showing no clear trend between the two parameters. However, when we consider only Non-RPS galaxies (blue dots) a trend emerges with increasing turnover radius for increasing bar length, specially for the case of the turnover radius identified using the star formation rate indicator. In the same plane, RPS galaxies (open circles) show a larger scatter and do not seem to follow a systematic trend, however, on average, low mass galaxies (log$(M_*/M_{\odot}) < 10)$) tend to have a shorter turnover radius  than high-mass ones (log$(M_*/M_{\odot}) \geq 10)$.

\subsection{Specific star formation profiles}
The star formation rate in galaxies is generally dependent on the stellar mass, both on a global and on a local scale \citep{Cano-Diaz19}. Another effective and complementary way to explore if and how stellar bars influence, together with ram pressure, the star formation processes, is to perform an analysis similar to the one performed in \autoref{sec:Central_Star_Formation}, removing the somehow subjective detection of the turnover radius. To this aim, we calculate average radial profiles of the specific star formation rates (sSFR) and, other than separating barred and unbarred galaxies, we also divide galaxies based on their ram pressure stage, something which should allow us to at least partially separate the effect of the bar and that of ram pressure. 

This analysis is presented in \autoref{specific}, where we show radial sSFR profiles (radial distances are normalized to the effective radius of each galaxy) for different sub-samples drawn from the main one. We have estimated the uncertainties on the sSFR for each subsample and at each normalized distance, using the bootstrap resampling, as described in Section \ref{sec:Central_Star_Formation}. In this way, the error bars for each point in Figure \ref{specific}, denote the estimated 1-sigma confidence intervals.

In the top--left panel, no differences in the profile are found when a simple distinction between barred (shown in red) and unbarred galaxies (in blue) is made. If instead we divide by RPS and Non RPS, only a marginal difference, within the dispersion, is found in the innermost part of the galaxies: on average, RPS galaxies show a mild tendency to have slightly higher sSFR values. 

Further separating the sample, and looking only at Non-RPS galaxies, again no differences are found in the sSFR profiles with respect to the presence/absence of a stellar bar (bottom--left panel). If instead we look at RPS galaxies at the peak stripping (jellyfish), segregating between barred and unbarred, we do find a highly significant difference in the sSFR in their very center, with barred galaxies displaying higher values of the sSFR by more than one order of magnitude. This is an indication that for the galaxies in our sample, bars are not sufficient to produce an upturn in the sSFR profiles, but when acting on strongly ram-pressure affected galaxies it produces the most dramatic changes on the profile, substantially increasing the central SFR.    

When contrasting ram-pressure affected with non-ram pressure affected galaxies (top--right panel), we found a minor difference on their sSFR profiles that we cannot attribute entirely to the ram-pressure effect, as bars most certainly are also contributing to this effect. Examining the sSFR profiles for the galaxies at different stages of interaction with the cluster gas, we only found a striking difference between barred and unbarred galaxies for the jellyfish sub-sample, reinforcing the idea that to produce such upturn we require both, the redistribution of gas by extreme ram-pressure and the fueling triggered by the presence of the bar.
  
Exploring the presence of AGNs in our sample, we note that all of them but two are located in barred galaxies. We propose the following scenario as a possibility to explain this occurrence. RP-peak galaxies that do not host a bar, have been already gone through the most intense stripping phase, and star formation has been severely affected, even in their centers. In fact, these are the objects for which the central sSFR is the lowest in the sample (see the bottom--right panel of \autoref{specific}). On the other side, the presence of a bar seems to not only mitigate the stripping effect, but also to promote central star formation. In fact, it has been argued observationally \citep{Poggianti++17} and found in numerical simulations \citep{Ramos18} that ram pressure is able to convey gas towards the center of galaxies, in this context the bar might serve not only as a funneling structure, but also as a gas tank. The idea here, is that as the bar sweeps the galaxy disk, it also fuels gas with the process becoming even more efficient when RP is acting, as it redistributes gas within the galaxy supplying extra fuel to the area affected by the bar. This gas takes about 250 Myr \citep{Carles16} to reach the central parts where it can boost both AGN and star formation activity. This is obliviously a transient phenomenon, that quickly stops as the gas in the bar is exhausted. In this instance it is quite interesting to note that the only peak-stripping galaxy that does not host an AGN is JO175 \citep{Poggianti++17}, for which we do not encounter evidences for the presence of a bar. While one object cannot indeed be considered as a final proof, this is coherent to the scenario proposed above. 

\section{SUMMARY \& CONCLUSIONS} \label{sec:conclusions}
According to galaxy formation theories in an inside-out scenario, the oldest stellar populations are expected, on average, to be found in the central part of galaxies, with a gradient towards younger populations to the outskirts. Hence, finding young populations in the innermost region of a galaxy could possibly suggest the presence of a star formation promoter mechanism in this region. In this work we have studied the influence of ram pressure together with the presence of a stellar bar as a mechanism affecting the star formation activity and the stellar populations in the central region of galaxies. 

To investigate the combined effect of bars in galaxies affected by ram-pressure stripping, we have used the GASP sample of 114 galaxies, an ambitious ESO program aimed at studying gas removal processes in nearby galaxies in different environments at $z=$ 0.04$-$0.07. In order to work with a reliable selection of barred and unbarred galaxies, we selected only face-on galaxies from the sample computing the minor-to-major axis ratio ($b/a>0.5$) using the external isophotes of the disk employing $i$-band images. We then identified the presence (or absence) of a stellar bar in each galaxy through visual inspection and a combination of position angle and eccentricity profiles of the isophotes. Finally, we removed galaxies that presented clear signatures of interactions with nearby companions, where additional processes taking place, and their possible combined effects are outside the scope of our analysis.

We produced two-dimensional maps of the star formation rate surface density, $\Sigma_{SFR}$, of the equivalent width of the H$\alpha$ line and of the luminosity weighted stellar age. From two-dimensional maps of these quantities, we proceeded to estimate the one-dimensional radial profiles of the same parameters. Then, in order to investigate the combined impact of bars and ram-pressure stripping, we identify a sudden change of the radial profiles within the central region of the galaxies ($<R_e$) by visual inspection, a point that we refer to as the turnover radius, $r_t$. For each SFH indicator we quantified its turnover strength by fitting a straight line to the points that were beyond $r_t$, as long as the uncertainty allowed. We compared the value of the observed profile at $r=0$ with the value of best-fitting line extrapolated at $r=0$, looking for a possible relationship between the radius and the strength of the turnover with the length of the bar.

Finally, in order to remove the somehow subjective detection of the turnover radius, we compute and analyze the sSFR profiles separating barred and unbarred galaxies at different stages of ram-pressure.  

Our main results can be summarized as follows: 

\begin{enumerate} 

    \item Galaxies that are going through ram pressure have a greater increase in star formation activity in the central region compared to galaxies that are not affected by this process.
    
    \item The most extreme cases of increase central star formation activity and rejuvenation of the stellar population that we find are in barred galaxies that are at the peak of ram pressure.
    
    \item When we segregate our sample by stellar mass, low-mass barred galaxies affected by ram pressure display a systematic increase in the SFR and a rejuvenation in the central region, while both enhancement and suppression are found for higher mass galaxies.
    
    \item For the case of Non-RPS galaxies, we find a correlation between the extent of the turnover radius and the length of the bar, especially when the turnover radius is identified through the inspection of the $\mathrm{\Sigma_{SFR}} $ profiles.
    
\end{enumerate}

An enhancement of the star formation activity in the central region of barred galaxies, as compared to unbarred ones, is a potential observational indication that the gas can more efficiently flow towards the center of the galaxy due to the presence of the bar. In fact, evidence for a short-lived enhancement of star formation in barred galaxies has been already reported \citep{Allard06, Ellison+11}. However, we expect the episodes of enhanced star formation in barred RPS galaxies to last longer and/or be more intense, as in these systems the ram pressure exerted on the cold interstellar medium by the hot intracluster medium can efficiently compress and redistribute it, funnelling gas to the bar, that otherwise would be out of its reach. Our results support this scenario, showing that the interaction between gas and the stellar bar is more evident in galaxies at the peak phase of ram-pressure stripping (jellyfish), as it would be expected given that enhanced star formation is already present in the disks of galaxies at this phase \citep{Vulcani18}. In this sense, ram pressure would act as an external mechanism to feed cold gas to the bar that otherwise would be unavailable for the bar to produce star formation, in contraposition to internal mechanisms such as strong spiral arm coupled to the corotation radius of the bar \citep{Masset97,Wang20}.

That ram pressure plays an important role in determining the efficiency/capability of forming stars inside bulges, is also highlighted by the comparison with barred galaxies that do not display evidences of ram-pressure interactions. While such a comparison cannot be made with a statistical significance for our sample, \cite{Ellison+11} find that (unperturbed) low-mass barred galaxies show no significant increase in their central SFR when contrasted with a control sample of unbarred galaxies. This is not seen when we look at the low-mass subsample of barred galaxies with signatures of ram pressure interactions, which instead presents a systematic increase in the central star formation activity and a consequent rejuvenation of the stellar populations as well.

An explanation of this result can be given by taking into account the fact that low-mass galaxies tend to have a proportionally larger reservoir of gas \citep{Masters12}, while at the same time presenting shorter bars \citep{Erwin19} as compared to high-mass galaxies. Hence, in the case of the former, the episodes of star formation activity in the central region can be of longer duration, although of much lower intensity exactly because of the weaker interaction between the gas disturbed by the RPS with the shorter bar itself.

In case of high-mass barred galaxies, the strength of both the SFR and the average stellar age shows a much larger spread in values, and both suppression and enhancement of star formation are found in the central region. This could be due to the presence of a larger bar that would induce shorter star formation activity periods but, at the same time, more intense, hence more rapidly depleting the gas reservoirs \citep{Friedli95, Martel18}. Furthermore, jellyfishes at the peak of ram pressure in the GASP sample, are mostly high-mass galaxies, for which the gas has been already strongly stripped. Observing such differences in the central star formation activity might hence reflect the observation of different phases of this process.

There is however another aspect that should be taken into account: nuclear activity. In fact, all galaxies with centrally suppressed star formation activity in the RPS sub-sample show indeed AGN activity, which could point to a suppression by this mechanism that might be linked or not to the presence of a bar, as the bar in principle can actively fuel gas to feed the central supermassive black hole. 

Recent observational evidence in this direction has been reported by \cite{Silva22}, who found that AGNs are preferentially found in barred galaxies, but also that the accretion rate is higher in barred galaxies. Furthermore, simulations also support this scenario. \cite{Rosas-Guevara20}, for instance,  found that the median mass of SMBHs hosted by the strongly barred galaxies at high redshift is systematically higher than the one hosted by unbarred galaxies.

 A possible relation between RP and the presence of AGN activity was explored in \cite{Poggianti++17}. Even though with a very small sample, clear evidence of nuclear activity was reported in 6 out of 7 galaxies with strong signs of RP. Due to the high inclination of the disk, only for 3 of them we were able to perform a photometric analysis to detect a stellar bar: 2 of them turned to be barred, while for one (JO175) we do not detect any signs of a bar. Sanchez-Garcia et al. (in prep) and Bacchini et al. (subm), using a dynamical stellar model, were able to clearly detect stellar bars in the remaining 3 (JO204, JO206 and JW100). AGN activity was indeed detected in all of them, apart from JO175, the only non-barred galaxy. Now, this is surely a small and non complete sample, and this interesting issue surely deserves more investigation, to check for the general ubiquity of the phenomenon and to better explore this relation. If the sample of strongly RP-affected galaxies is extended, and AGN activity is also explored by other means (i.e. such as X-ray luminosity, as done in \citealt{Peluso22}), then we do find AGN activity also in non-barred galaxies.

Another aspect we have explored, is the correlation between the turnover radius and the bar length also reported by \cite{Lin17} and \cite{Lin20}. They found that the larger the length of the bar, the larger the radius of the turnover. For the case of Non-RPS galaxies, we find a similar trend --even though weaker than previously found-- when the identification of the turnover radius is performed using the $\mathrm{\Sigma_{SFR}} $SFR profiles. The fact that we have not found a relationship as tight as \cite{Lin17} and \cite{Lin20} can be attributed to several factors, among them subjectivity in the identification of the turnover radius, the area the spatial coverage of the galaxy disk, the resolution and the stochastic nature of the different samples involved. For the case of the galaxies that are experiencing the effect of RP, we do not find any correlation. We attribute this to the added effect of ram-pressure, since this mechanism, depending both on the stage in which it is observed and on the particular configuration of the galaxy, can induce dramatic modifications on the distribution of the gas and the induce star formation activity.

The morphological transformation to which galaxies are subject to in the cluster environment, are relatively fast, and evidence is mounting suggesting that they end up disrupting stellar bars, when present \citep{Tawfeek22}. But our work has shown that, when a gas-rich barred spiral enters a cluster, the combination of ram pressure with its presence, is a quite efficient mechanism to funnel gas towards the center, hence favoring the onset of central star formation activity and, most likely, enhancing the probability of nuclear activity \citep{Poggianti++17}.

\begin{acknowledgments} The authors thank the referee for constructive comments that helped to improve and clarify the manuscript. Based on observations collected at the European Organization for Astronomical Research in the Southern Hemisphere under ESO programme 196.B-0578. This project has received funding from the European Research Council (ERC) under the European Union's Horizon 2020 research and innovation programme (grant agreement No. 833824). J.F. and O.S. acknowledge financial support from the UNAM-DGAPA-PAPIIT IN111620 grant, M\'exico. B.C.S. and O.S. acknowledge financial support through PAPIIT projects IA103520 and IN108323, from DGAPA-UNAM, M\'exico. A.M., B.M.P., M.G. and B.V acknowledge funding from the INAF main-stream funding programme (PI B. Vulcani). B.V. and M.G. acknowledge the Italian PRIN-Miur 2017 (PI A. Cimatti).
\end{acknowledgments}

\newpage
\appendix

\section{Appendix information}

\startlongtable
\begin{deluxetable*}{lccccccccl}
\tablenum{2}
\tablecaption{Properties of the galaxies in our sample\label{tab:general}}
\tablewidth{0pt}
\tablehead{
\colhead{Galaxy name} &
\colhead{ Bar}  & \colhead{Bar length}& \colhead{Bar length} & \colhead{P.A. bar }& \colhead{log $M_*$}& \colhead{b/a} & \colhead{$r_e$} &  \colhead{$r_e$} & \colhead{Stripping stage}\\
 & \colhead{1/0}  & \colhead{(arcsec)}& \colhead{(kpc)} & \colhead{(degrees)}& \colhead{$M_{\odot}$}& \colhead{(galaxy)} & \colhead{(arcsec)} & \colhead{(kpc)} &
}
\decimalcolnumbers
\startdata
JO5            &  0  &   --   &   --   &    --    &  10.3 & 0.747 & 3.833  & 4.784 & non\_stripped \\
JO10           &  0  &   --   &   --   &    --    &  10.8 & 0.508 & 5.668  & 5.266 & truncated \\
JO13           &  0  &   --   &   --   &    --    &  9.8  & 0.969 & 4.200  & 3.948 & initial \\
JO17           &  0  &   --   &   --   &    --    &  10.2 & 0.525 & 6.035  & 5.353 & initial \\
JO41           &  1  &  2.288 &  2.146 &  21.339  &  10.2 & 0.867 & 4.567  & 4.284 & initial \\
JO47           &  1  &  2.067 &  1.753 &  5.643   &  9.6  & 0.915 & 4.934  & 4.184 & stripped \\
JO49           &  1  &  4.405 &  3.903 &  171.411 &  10.7 & 0.640 & 6.402  & 5.672 & jelly \\
JO68           &  1  &  3.555 &  3.878 &  161.315 &  10.0 & 0.523 & 4.567  & 4.983 & initial \\
JO69           &  1  &  2.207 &  2.408 &  74.358  &  9.9  & 0.755 & 4.567  & 4.983 & stripped \\
JO70           &  1  &  3.144 &  3.430 &  98.714  &  10.5 & 0.799 & 3.466  & 3.781 & stripped \\
JO73           &  1  &  2.322 &  3.158 &  67.465  &  10.0 & 0.880 & 4.200  & 5.712 & initial \\
JO85           &  1  &  6.346 &  4.518 &  80.225  &  10.7 & 0.866 & 7.136  & 5.081 & stripped \\
JO93           &  0  &   --   &   --   &    --    &  10.5 & 0.932 & 7.503  & 5.515 & initial \\
JO95           &  1  &  3.019 &  2.575 &  7.786   &  9.3  & 0.861 & 4.567  & 3.896 & jelly \\
JO112          &  0  &   --   &   --   &    --    &  9.6  & 0.891 & 2.732  & 3.079 & initial \\
JO119          &  0  &   --   &   --   &    --    &  10.4 & 0.952 & 4.200  & 4.066 & initial \\
JO123          &  1  &  2.457 &  2.629 &  36.249  &  9.9  & 0.872 & 3.466  & 3.709 & initial \\
JO128          &  1  &  2.624 &  2.571 &  1.266   &  9.9  & 0.965 & 4.934  & 4.835 & initial \\
JO153          &  0  &   --   &   --   &    --    &  9.5  & 0.944 & 3.099  & 2.863 & stripped \\
JO156          &  0  &   --   &   --   &    --    &  9.6  & 0.571 & 5.301  & 5.301 & stripped \\
JO157          &  1  &  3.232 &  2.854 &  126.377 &  10.1 & 0.919 & 4.567  & 4.033 & stripped \\
JO159          &  1  &  1.590 &  1.496 &  118.925 &  9.8  & 0.850 & 3.466  & 3.262 & initial \\
JO160          &  1  &  1.653 &  1.564 &  172.428 &  10.1 & 0.513 & 5.301  & 5.015 & jelly \\
JO171          &  0  &   --   &   --   &    --    &  10.6 & 0.891 & 9.338  & 9.469 & jelly \\
JO175          &  0  &   --   &   --   &    --    &  10.5 & 0.845 & 3.099  & 2.845 & jelly \\
JO179          &  1  &  2.046 &  2.441 &  177.712 &  9.5  & 0.745 & 3.466  & 4.135 & initial \\
JO180          &  0  &   --   &   --   &    --    &  10.0 & 0.971 & 3.466  & 4.312 & stripped \\
JO181          &  0  &   --   &   --   &    --    &  9.3  & 0.513 & 2.732  & 3.153 & initial \\
JO194          &  1  &  5.087 &  4.217 &  104.06  &  11.2 & 0.745 & 10.072 & 8.350 & jelly \\
JO197          &  0  &   --   &   --   &    --    &  10.0 & 0.535 & 4.200  & 4.586 & stripped \\
JO200          &  1  &  2.156 &  2.221 &  99.349  &  10.8 & 0.603 & 9.338  & 9.618 & stripped \\
JO201          &  1  &  5.345 &  4.613 &  87.563  &  10.9 & 0.640 & 6.769  & 5.943 & jelly \\
JO205          &  0  &   --   &   --   &    --    &  9.5  & 0.698 & 3.833  & 3.377 & initial \\
JW10           &  1  &  3.911 &  5.358 &  138.202 &  10.0 & 0.785 & 4.200  & 5.754 & initial \\
JW39           &  0  &   --   &   --   &    --    &  11.2 & 0.849 & 6.769  & 8.597 & jelly \\
JW105          &  1  &  2.883 &  2.906 &  1.792   &  10.1 & 0.928 & 1.998  & 2.014 & non\_stripped \\
P443           &  1  &  5.729 &  5.207 &  77.354  &  10.7 & 0.840 & 2.732  & 2.483 & non\_stripped \\
P648           &  0  &   --   &   --   &    --    &  10.4 & 0.641 & 5.668  & 7.187 & non\_stripped \\
P669           &  0  &   --   &   --   &    --    &  10.5 & 0.645 & 7.136  & 6.401 & non\_stripped \\
P954           &  1  &  2.137 &  1.895 &  129.82  &  9.6  & 0.649 & 4.200  & 3.725 & non\_stripped \\
P5169          &  0  &   --   &   --   &    --    &  10.6 & 0.615 & 2.365  & 2.892 & non\_stripped \\
P5215          &  0  &   --   &   --   &    --    &  10.5 & 0.708 & 5.301  & 6.435 & non\_stripped \\
P11695         &  0  &   --   &   --   &    --    &  9.9  & 0.882 & 4.934  & 4.500 & non\_stripped \\
P13384         &  0  &   --   &   --   &    --    &  9.8  & 0.893 & 3.466  & 3.463 & non\_stripped \\
P14672         &  0  &   --   &   --   &    --    &  9.8  & 0.884 & 3.099  & 3.018 & non\_stripped \\
P16762         &  0  &   --   &   --   &    --    &  10.7 & 0.518 & 3.833  & 3.649 & non\_stripped \\
P17048         &  1  &  1.100 &  1.054 &  164.208 &  9.5  & 0.879 & 3.099  & 2.969 & non\_stripped \\
P17945         &  0  &   --   &   --   &    --    &  9.7  & 0.674 & 4.200  & 3.629 & non\_stripped \\
P18060         &  0  &   --   &   --   &    --    &  8.9  & 0.572 & 2.732  & 2.322 & non\_stripped \\
P20769         &  0  &   --   &   --   &    --    &  9.4  & 0.724 & 2.365  & 2.263 & non\_stripped \\
P20883         &  0  &   --   &   --   &    --    &  9.8  & 0.582 & 4.200  & 4.973 & non\_stripped \\
P21734         &  1  &  3.071 &  4.029 &  172.958 &  10.8 & 0.785 & 6.035  & 7.918 & non\_stripped \\
P25500         &  0  &   --   &   --   &    --    &  10.8 & 0.625 & 8.604  & 10.024 & non\_stripped \\
P42932         &  0  &   --   &   --   &    --    &  10.5 & 0.678 & 6.402  & 5.186 & non\_stripped \\
P45479         &  1  &  2.724 &  2.737 &  123.558 &  10.6 & 0.643 & 5.301  & 5.328 & non\_stripped \\
P48157         &  1  &  3.753 &  4.443 &  8.726   &  10.5 & 0.720 & 6.035  & 7.145 & non\_stripped \\
P63661         &  0  &   --   &   --   &    --    &  10.2 & 0.607 & 6.402  & 6.863 & non\_stripped \\
P63947         &  1  &  1.384 &  1.510 &  71.287  &  9.3  & 0.908 & 2.365  & 2.580 & non\_stripped \\
P95080         &  1  &  1.043 &  0.832 &  72.043  &  9.9  & 0.686 & 7.136  & 5.695 & non\_stripped \\
P96244         &  1  &  2.736 &  2.826 &  63.105  &  10.7 & 0.561 & 6.402  & 6.613 & stripping \\
A500\_F\_0152  &  0  &   --   &   --   &    --    &  9.2  & 0.584 & 1.998  & 2.661 & non\_stripped \\
A3128\_B\_0148 &  0  &   --   &   --   &    --    &  9.8  & 0.618 & 2.365  & 2.635 & non\_stripped \\
A3158\_11\_91  &  1  &  2.030 &  2.393 &  160.698 &  10.1 & 0.751 & 4.200  & 4.952 & non\_stripped \\
A3158\_B\_0223 &  1  &  3.841 &  4.229 &  110.242 &  10.3 & 0.974 & 3.099  & 3.412 & non\_stripped \\
A3158\_B\_0234 &  0  &   --   &   --   &    --    &  9.9  & 0.601 & 2.732  & 3.461 & non\_stripped \\
A3376\_B\_0261 &  0  &   --   &   --   &    --    &  10.5 & 0.991 & 4.200  & 4.150 & non\_stripped
\enddata 
\tablecomments{Note:\\
(1) Galaxy name.\\
(2) Bar. 1:Barred galaxy, 0:Unbarred galaxy\\
(3) Bar length (arsec) \\
(4) Position angle of the bar (degrees)\\
(5) Galaxy mass (log$(M_*/M_{\odot}$)\\
(6) Minor-to-major axial ratio (b/a) of the galaxy.\\
(7) Effective Radius (arcsec). \\
(8) Stripping stage.
}
\end{deluxetable*}
\onecolumngrid


\begin{thebibliography}{}

\bibitem[\protect\citeauthoryear{Allard et al.}{2006}]{Allard06} Allard, E.~L., Knapen, J.~H., Peletier, R.~F., Sarzi, M., 2006, MNRAS, 371, 1087
\bibitem[Athanassoula(1992)]{Athanassoula92} Athanassoula, E.\ 1992, \mnras, 259, 345 
\bibitem[Athanassoula(2005)]{Athanassoula05} Athanassoula, E.\ 2005, \mnras, 358, 1477
\bibitem[Athanassoula \& Misiriotis(2002)]{Athana02} Athanassoula, E., Misiriotis, A., 2002, MNRAS, 330, 35
\bibitem[\protect\citeauthoryear{Athanassoula}{2003}]{Athanassoula03} Athanassoula, E., 2003, MNRAS, 341, 1179
\bibitem[\protect\citeauthoryear{Athanassoula}{2013}]{Athanassoula13} Athanassoula, E.\ 2013, Secular Evolution of Galaxies, 305
\bibitem[\protect\citeauthoryear{Aguerri, M{\'e}ndez-Abreu, \& Corsini}{2009}]{Aguerri09} Aguerri, J.~A.~L., M{\'e}ndez-Abreu, J., Corsini, E.~M., 2009, A\&A, 495, 491

\bibitem[Bacon et al.(2010)]{Bacon10} Bacon, R., Accardo, M., Adjali, L., et al.\ 2010, \procspie, 773508
\bibitem[Balogh et al.(2000)]{Balogh00} Balogh, M.~L., Navarro, J.~F., \& Morris, S.~L.\ 2000, \apj, 540, 113
\bibitem[\protect\citeauthoryear{Baldwin, Phillips \& Terlevich}{1981}]{Baldwin81} Baldwin, J.~A., Phillips, M.~M., Terlevich, R., 1981, PASP, 93, 5
\bibitem[\protect\citeauthoryear{Barazza, Jogee \& Marinova}{2008}]{Barazza08} Barazza, F.~D., Jogee, S., Marinova, I., 2008, ApJ, 675, 1194
\bibitem[Bekki (2014)]{Bekki14} Bekki K., 2014, MNRAS, 438, 444
\bibitem[Bekki (2003)]{Bekki03} Bekki K., Couch W.~J., 2003, ApJL, 596, L13
\bibitem[Berentzen et al.(1998)]{Berentzen98} Berentzen, I., Heller, C.~H., Shlosman, I., et al.\ 1998, \mnras, 300, 49
\bibitem[Berentzen et al.(2007)]{Berentzen07} Berentzen, I., Shlosman, I., Martinez-Valpuesta, I., \& Heller, C.~H.\ 2007, \apj, 666, 189 
\bibitem[Bitsakis et al.(2019)]{Bitsakis19} Bitsakis, T., S{\'a}nchez, S.~F., Ciesla, L., et al.\ 2019, \mnras, 483, 370
\bibitem[\protect\citeauthoryear{Boselli et al.}{2019}]{Boselli19} Boselli, A., Epinat, B., Contini, T., Abril-Melgarejo, V., Boogaard, L.~A., Pointecouteau, E., Ventou, E., et al., 2019, A\&A, 631, A114
\bibitem[\protect\citeauthoryear{Bournaud \& Combes}{2002}]{Bournaud02} Bournaud, F., Combes, F., 2002, A\&A, 392, 83
\bibitem[\protect\citeauthoryear{Bournaud, Combes, \& Semelin}{2005}]{Bournaud05} Bournaud, F., Combes, F., Semelin, B., 2005, MNRAS, 364, L18
\bibitem[Buta et al.(2010)]{Buta10} Buta, R.~J., Sheth, K., Regan, M., et al.\ 2010, \apjs, 190, 147
\bibitem[Byrd \& Valtonen(1990)]{Byrd90} Byrd, G., \& Valtonen, M.\ 1990, \apj, 350, 89

\bibitem[Calvi et al.(2011)]{calvi11} Calvi, R., Poggianti, B.~M., \& Vulcani, B.\ 2011, \mnras, 416, 727
\bibitem[\protect\citeauthoryear{Cano-D{\'\i}az et al.}{2019}]{Cano-Diaz19} Cano-D{\'\i}az, M., {\'A}vila-Reese, V., S{\'a}nchez, S.~F., Hern{\'a}ndez-Toledo, H.~M., Rodr{\'\i}guez-Puebla, A., Boquien, M., Ibarra-Medel, H., 2019, MNRAS, 488, 3929
\bibitem[\protect\citeauthoryear{Cappellari \& Emsellem}{2004}]{Cappellari04} Cappellari, M., Emsellem, E., 2004, PASP, 116, 138
\bibitem[\protect\citeauthoryear{Cardelli, Clayton \& Mathis}{1989}]{Cardelli89} Cardelli, J.~A., Clayton, G.~C., Mathis, J.~S., 1989, ApJ, 345, 245
\bibitem[\protect\citeauthoryear{Carles et al.}{2016}]{Carles16} Carles, C., Martel, H., Ellison, S.~L., Kawata, D., 2016, MNRAS, 463, 1074.
\bibitem[\protect\citeauthoryear{Catal{\'a}n-Torrecilla, et al.}{2017}]{Catalan+17} Catal{\'a}n-Torrecilla C., et al., 2017, ApJ, 848, 87
\bibitem[\protect\citeauthoryear{Catinella et al.}{2010}]{Catinella10} Catinella, B., Schiminovich, D., Kauffmann, G., Fabello, S., Wang, J., Hummels, C., Lemonias, J., et al., 2010, MNRAS, 403, 683. 
\bibitem[Cervantes Sodi et al.(2015)]{CervantesSodi15} Cervantes Sodi, B., Li, C., \& Park, C.\ 2015, \apj, 807, 111
\bibitem[Cervantes Sodi(2017)]{CervantesSodi17} Cervantes Sodi, B.\ 2017, \apj, 835, 80
\bibitem[\protect\citeauthoryear{Cervantes Sodi \& S{\'a}nchez Garc{\'\i}a}{2017}]{Cervantes-Sanchez17} Cervantes Sodi, B., S{\'a}nchez Garc{\'\i}a, O., 2017, ApJ, 847, 37
\bibitem[Chabrier(2003)]{chabrier03} Chabrier, G.\ 2003, \pasp, 115, 763
\bibitem[Chown et al.(2019)]{Chown19} Chown, R., Li, C., Athanassoula, E., et al.\ 2019, \mnras, 484, 5192
\bibitem[\protect\citeauthoryear{Cid Fernandes et al.}{2011}]{Cid11} Cid Fernandes R., Stasi{\'n}ska G., Mateus A., Vale Asari N., 2011, MNRAS, 413, 1687
\bibitem[\protect\citeauthoryear{Cid Fernandes et al.}{2010}]{Cid10} Cid Fernandes R., Stasi{\'n}ska G., Schlickmann M.~S., Mateus A., Vale Asari N., Schoenell W., Sodr{\'e} L., 2010, MNRAS, 403, 1036
\bibitem[Coelho \& Gadotti(2011)]{Coelho11} Coelho, P., \& Gadotti, D.~A.\ 2011, \apjl, 743, L13
\bibitem[Combes \& Elmegreen(1993)]{Combes93} Combes, F., \& Elmegreen, B.~G.\ 1993, \aap, 271, 391
\bibitem[Combes \& Gerin(1985)]{Combes85} Combes, F., \& Gerin, M.\ 1985, \aap, 150, 327
\bibitem[\protect\citeauthoryear{Consolandi, et al.}{2017}]{Consolandi+17} Consolandi G., Dotti M., Boselli A., Gavazzi G., Gargiulo F., 2017, A\&A, 598, A114
\bibitem[Cowie \& Songaila(1977)]{Cowie77} Cowie, L.~L., \& Songaila, A.\ 1977, \nat, 266, 501

\bibitem[Deb et al.(2020)]{deb20} Deb, T., Verheijen, M.~A.~W., Gullieuszik, M., et al.\ 2020, \mnras, 494, 5029
\bibitem[de Vaucouleurs et al.(1991)]{deVaucouleurs91} de Vaucouleurs, G., de Vaucouleurs, A., Corwin, H. G. Jr., et al. 1991, Third
Reference Catalogue of Bright Galaxies (New York: Springer)

\bibitem[\protect\citeauthoryear{Ellison et al.}{2011}]{Ellison+11} Ellison S.~L., Nair, P., Patton, D.~R., Scudder, J.~M., Mendel, J.~T., Simard, L., 2011, MNRAS, 416, 2182
\bibitem[\protect\citeauthoryear{Erwin}{2019}]{Erwin19} Erwin, P., 2019, MNRAS, 489, 3553
\bibitem[\protect\citeauthoryear{Eskridge, et al.}{2000}]{Eskridge00} Eskridge, P.~B., et al., 2000, AJ, 119, 536

\bibitem[Fasano et al.(2006)]{fasano06} Fasano, G., Marmo, C., Varela, J., et al.\ 2006, \aap, 445, 805
\bibitem[Ferland et al.(1998)]{ferland98} Ferland, G.~J., Korista, K.~T., Verner, D.~A., et al.\ 1998, \pasp, 110, 761
\bibitem[Ferland et al.(2013)]{ferland13} Ferland, G.~J., Porter, R.~L., van Hoof, P.~A.~M., et al.\ 2013, \rmxaa, 49, 137
\bibitem[\protect\citeauthoryear{Fossati, et al.}{2016}]{Fossati16} Fossati, M., Fumagalli, M., Boselli, A., Gavazzi, G., Sun, M., Wilman, D.~J., 2016, MNRAS, 455, 2028

\bibitem[\protect\citeauthoryear{Friedli \& Benz}{1995}]{Friedli95} Friedli, D., Benz, W., 1995, A\&A, 301, 649

\bibitem[Fritz et al.(2007)]{fritz07} Fritz, J., Poggianti, B.~M., Bettoni, D., et al.\ 2007, \aap, 470, 137
\bibitem[Fritz et al.(2017)]{fritz17} Fritz, J., Moretti, A., Gullieuszik, M., et al.\ 2017, \apj, 848, 132

\bibitem[\protect\citeauthoryear{George et al.}{2019}]{George19} George, K., Joseph, P., Mondal, C., Subramanian, S., Subramaniam, A., Paul, K.~T., 2019, A\&A, 621, L4

\bibitem[\protect\citeauthoryear{G{\'e}ron et al.}{2021}]{Geron21} G{\'e}ron, T., Smethurst, R.~J., Lintott, C., Kruk, S., Masters, K.~L., Simmons, B., Stark, D.~V., 2021, MNRAS, 507, 4389
\bibitem[Gullieuszik et al.(2015)]{gullieuszik15} Gullieuszik, M., Poggianti, B., Fasano, G., et al.\ 2015, \aap, 581, A41
\bibitem[Gullieuszik et al.(2020)]{Gullieuszik20} Gullieuszik, M., Poggianti, B.~M., McGee, S.~L., et al.\ 2020, \apj, 899, 13

\bibitem[Gunn \& Gott(1972)]{GunnGott72} Gunn, J.~E., \& Gott, J.~R.\ 1972, \apj, 176, 1

\bibitem[\protect\citeauthoryear{Huang et al.}{2012}]{Huang12} Huang, S., Haynes, M.~P., Giovanelli, R., Brinchmann, J., 2012, ApJ, 756, 113
\bibitem[Hunt et al.(2008)]{Hunt08} Hunt, L.~K., Combes, F., Garc{\'\i}a-Burillo, S., et al.\ 2008, \aap, 482, 133

\bibitem[Kapferer et al.(2009)]{Kapferer09} Kapferer W., Sluka C., Schindler S., Ferrari C., Ziegler B., 2009, A\&A, 499, 87
\bibitem[\protect\citeauthoryear{Kauffmann, et al.}{2003}]{Kauffmann03} Kauffmann, G., et al., 2003, MNRAS, 346, 1055
\bibitem[\protect\citeauthoryear{Kennicutt}{1998}]{Kennicutt98} Kennicutt, R.~C., 1998, ARA\&A, 36, 189
\bibitem[\protect\citeauthoryear{Kewley, et al.}{2001}]{Kewley01} Kewley, L.~J., Heisler, C.~A., Dopita, M.~A., Lumsden, S., 2001, ApJS, 132, 37

\bibitem[Kim et al.(2017)]{Kim17} Kim, E., Hwang, H.~S., Chung, H., et al.\ 2017, \apj, 845, 93
\bibitem[Kormendy \& Kennicutt(2004)]{Kormendy04} Kormendy, J., \& Kennicutt, R.~C., Jr.\ 2004, \araa, 42, 603 
\bibitem[Kronberger et al.(2008)]{Kronberger08} Kronberger T., Kapferer W., Ferrari C., Unterguggenberger S., Schindler S., 2008, A\&A, 481, 337
\bibitem[\protect\citeauthoryear{Krumholz \& Kruijssen}{2015}]{Krumholz15} Krumholz, M.~R., Kruijssen, J.~M.~D., 2015, MNRAS, 453, 739
\bibitem[Kubryk et al.(2015)]{Kubryk15} Kubryk, M., Prantzos, N., \& Athanassoula, E.\ 2015, \aap, 580, A127

\bibitem[Larson et al.(1980)]{Larson80} Larson, R.~B., Tinsley, B.~M., \& Caldwell, C.~N.\ 1980, \apj, 237, 692
\bibitem[Lee, et al.(2012)]{Lee+12} Lee, G.-H., Park, C., Lee, M.~G., Choi, Y.-Y., 2012, ApJ, 745, 125
\bibitem[\protect\citeauthoryear{Lee, Ann \& Park}{2019}]{Lee19} Lee, Y.~H., Ann, H.~B., Park, M.-G., 2019, ApJ, 872, 97
\bibitem[Lee et al. (2022)]{Lee22} Lee J.~H., Lee M.~G., Mun J.~Y., Cho B.~S., Kang J., 2022, ApJL, 931, L22
\bibitem[Lin, et al.(2010)]{Lin10} Lin, L., et al., 2010, ApJ, 718, 1158
\bibitem[Lin, et al.(2014)]{Lin14} Lin, Y., Cervantes Sodi B., Li C., Wang L., Wang E., 2014, ApJ, 796, 98
\bibitem[Lin et al.(2017)]{Lin17} Lin, L., Li, C., He, Y., et al.\ 2017, \apj, 838, 105
\bibitem[\protect\citeauthoryear{Lin et al.}{2020}]{Lin20} Lin, L., Li, C., Du, C., Wang, E., Xiao, T., Bureau, M., Fraser-McKelvie, A., et al., 2020, MNRAS, 499, 1406

\bibitem[\protect\citeauthoryear{Martel et al.}{2018}]{Martel18} Martel, H., Carles, C., Robichaud, F., Ellison, S.~L., Williamson, D.~J., 2018, MNRAS, 477, 5367
\bibitem[Masset \& Tagger(1997)]{Masset97} Masset F., Tagger M., 1997, A\&A, 322, 442
\bibitem[\protect\citeauthoryear{Masters et al.}{2012}]{Masters12} Masters, K.~L., Nichol, R.~C., Haynes, M.~P., Keel, W.~C., Lintott, C., Simmons, B., Skibba, R., et al., 2012, MNRAS, 424, 2180
\bibitem[\protect\citeauthoryear{Masters, et al.}{2011}]{Masters11} Masters, K.~L., et al., 2011, MNRAS, 411, 2026
\bibitem[\protect\citeauthoryear{McPartland, et al.}{2016}]{McPartland+16} McPartland, C., Ebeling, H., Roediger, E., Blumenthal, K., 2016, MNRAS, 455, 2994
\bibitem[\protect\citeauthoryear{Men{\'e}ndez-Delmestre, et al.}{2007}]{Menendez07} Men{\'e}ndez-Delmestre, K., Sheth, K., Schinnerer, E., Jarrett, T.~H., Scoville, N.~Z., 2007, ApJ, 657, 790
\bibitem[\protect\citeauthoryear{Merluzzi, et al.}{2016}]{Merluzzi+16} Merluzzi, P., Busarello, G., Dopita, M.~A., Haines, C.~P., Steinhauser, D., Bourdin, H., Mazzotta, P., 2016, MNRAS, 460, 3345
\bibitem[Mihos \& Hernquist(1994)]{Mihos94} Mihos, J.~C., \& Hernquist, L.\ 1994, \apjl, 425, L13
\bibitem[\protect\citeauthoryear{Miwa \& Noguchi}{1998}]{Miwa98} Miwa, T., Noguchi, M., 1998, ApJ, 499, 149
\bibitem[\protect\citeauthoryear{Moretti et al.}{2018}]{Moretti18} Moretti, A., Poggianti, B.~M., Gullieuszik, M., Mapelli, M., Jaff{\'e}, Y.~L., Fritz, J., Biviano, A., et al., 2018, MNRAS, 475, 4055
\bibitem[\protect\citeauthoryear{Moretti et al.}{2022}]{Moretti22} Moretti, A., Radovich, M., Poggianti, B.~M., Vulcani, B., Gullieuszik, M., Werle, A., Bellhouse, C., et al., 2022, ApJ, 925, 4
\bibitem[Moore et al.(1996)]{Moore96} Moore, B., Katz, N., Lake, G., et al.\ 1996, \nat, 379, 613
\bibitem[Mu{\~n}oz-Mateos et al.(2013)]{munoz-mateos13} Mu{\~n}oz-Mateos, J.~C., Sheth, K., Gil de Paz, A., et al.\ 2013, \apj, 771, 59


\bibitem[Nair \& Abraham(2010)]{Nair10} Nair, P.~B., Abraham, R.~G., 2010, ApJL, 714, L260
\bibitem[Neumann, et al.(2019)]{Neumann19} Neumann, J., et al., 2019, A\&A, 627, A26
\bibitem[\protect\citeauthoryear{Newnham}{2019}]{Newnham19} Newnham, L.~C., 2019, AAS
\bibitem[\protect\citeauthoryear{Noguchi}{1996}]{Noguchi96} Noguchi, M., 1996, ApJ, 469, 605


\bibitem[\protect\citeauthoryear{Osterbrock \& Ferland}{2006}]{Osterbrock06} Osterbrock, D.~E., Ferland, G.~J., 2006, agna.book

\bibitem[Park \& Choi(2009)]{Park09} Park, C., \& Choi, Y.-Y.\ 2009, \apj, 691, 1828
\bibitem[Peng, et al.(2002)]{peng02} Peng, C.~Y., Ho, L.~C., Impey, C.~D., Rix, H.-W., 2002, AJ, 124, 266
\bibitem[Peng et al.(2010)]{Peng10} Peng, Y.-. jie ., Lilly, S.~J., Kova{\v{c}}, K., et al.\ 2010, \apj, 721, 193
\bibitem[Peng et al.(2012)]{Peng12} Peng, Y.-. jie ., Lilly, S.~J., Renzini, A., et al.\ 2012, \apj, 757, 4
\bibitem[\protect\citeauthoryear{Peluso et al.}{2022}]{Peluso22} Peluso, G., Vulcani, B., Poggianti, B.~M., Moretti, A., Radovich, M., Smith, R., Jaff{\'e}, Y.~L., et al., 2022, ApJ, 927, 130
\bibitem[Poggianti et al.(2016)]{Poggianti+16} Poggianti, B.~M., Fasano, G., Omizzolo, A., et al.\ 2016, \aj, 151, 78
\bibitem[Poggianti et al.(2017)]{Poggianti17} Poggianti, B.~M., Moretti, A., Gullieuszik, M., et al.\ 2017, \apj, 844, 48
\bibitem[\protect\citeauthoryear{Poggianti et al.}{2017}]{Poggianti++17} Poggianti, B.~M., Jaff{\'e}, Y.~L., Moretti, A., Gullieuszik, M., Radovich, M., Tonnesen, S., Fritz, J., et al., 2017, Natur, 548, 304

\bibitem[\protect\citeauthoryear{Ramos-Mart{\'\i}nez, G{\'o}mez, \& P{\'e}rez-Villegas}{2018}]{Ramos18} Ramos-Mart{\'\i}nez, M., G{\'o}mez, G.~C., P{\'e}rez-Villegas, {\'A}., 2018, MNRAS, 476, 3781
\bibitem[\protect\citeauthoryear{Rosas-Guevara et al.}{2020}]{Rosas-Guevara20} Rosas-Guevara, Y., Bonoli, S., Dotti, M., Zana, T., Nelson, D., Pillepich, A., Ho, L.~C., et al., 2020, MNRAS, 491, 2547

\bibitem[Saintonge et al.(2011)]{Saintonge11} Saintonge, A., Kauffmann, G., Wang, J., et al.\ 2011, \mnras, 415, 61
\bibitem[\protect\citeauthoryear{Schlafly \& Finkbeiner}{2011}]{Schlafly11} Schlafly, E.~F., Finkbeiner, D.~P., 2011, ApJ, 737, 103
\bibitem[Sellwood \& Wilkinson(1993)]{Sellwood93} Sellwood, J.~A., \& Wilkinson, A.\ 1993, Reports on Progress in Physics, 56, 173 
\bibitem[\protect\citeauthoryear{Silva-Lima et al.}{2022}]{Silva22} Silva-Lima, L.~A., Martins, L.~P., Coelho, P.~R.~T., Gadotti, D.~A., 2022, arXiv, arXiv:2203.07794

\bibitem[\protect\citeauthoryear{Smith, et al.}{2010}]{Smith+10} Smith R.~J., et al., 2010, MNRAS, 408, 1417
\bibitem[\protect\citeauthoryear{Steinhauser et al.}{2012}]{Steinhauser12} Steinhauser, D., Haider, M., Kapferer, W., Schindler, S., 2012, A\&A, 544, A54
\bibitem[Steinhauser, Schindler, \& Springel (2016)]{Steinhauser16} Steinhauser D., Schindler S., Springel V., 2016, A\&A, 591, A51

\bibitem[\protect\citeauthoryear{Tawfeek et al.}{2022}]{Tawfeek22} Tawfeek A.~A., Cervantes Sodi B., Fritz J., Moretti A., P{\'e}rez-Mill{\'a}n D., Gullieuszik M., Poggianti B.~M., et al., 2022, ApJ, 940, 1

\bibitem[Toomre(1977)]{Toomre77} Toomre, A.\ 1977, Evolution of Galaxies and Stellar Populations, 401
\bibitem[\protect\citeauthoryear{Vulcani et al.}{2020}]{Vulcani20} Vulcani, B., Poggianti, B.~M., Tonnesen, S., McGee, S.~L., Moretti, A., Fritz, J., Gullieuszik, M., et al., 2020, ApJ, 899, 98
\bibitem[\protect\citeauthoryear{Vulcani et al.}{2021}]{Vulcani21} Vulcani, B., Poggianti, B.~M., Moretti, A., Franchetto, A., Bacchini, C., McGee, S., Jaff{\'e}, Y.~L., et al., 2021, ApJ, 914, 27

\bibitem[Valluri(1993)]{Valluri93} Valluri, M.\ 1993, \apj, 408, 57
\bibitem[\protect\citeauthoryear{Vulcani et al.}{2017}]{Vulcani17} Vulcani, B., Moretti, A., Poggianti, B.~M., Fasano, G., Fritz, J., Gullieuszik, M., Duc, P.-A., et al., 2017, ApJ, 850, 163



\bibitem[Vulcani et al.(2018)]{Vulcani18} Vulcani B., Poggianti B.~M., Gullieuszik M., Moretti A., Tonnesen S., Jaff{\'e} Y.~L., Fritz J., et al., 2018, ApJL, 866, L25


\bibitem[Wang et al. (2020)]{Wang20} Wang J., Athanassoula E., Yu S.-Y., Wolf C., Shao L., Gao H., Randriamampandry T.~H., 2020, ApJ, 893, 19
\bibitem[Wang et al.(2012)]{Wang12} Wang, J., Kauffmann, G., Overzier, R., et al.\ 2012, \mnras, 423, 3486
\bibitem[Wang et al.(2018)]{Wang18} Wang, E., Wang, H., Mo, H., et al.\ 2018, \apj, 860, 102
\bibitem[Weinberg(1985)]{Weinberg85} Weinberg, M.~D.\ 1985, \mnras, 213, 451 
\bibitem[Wozniak, et al.(1995)]{Wozniak95} Wozniak, H., Friedli, D., Martinet, L., Martin, P., Bratschi, P., 1995, A\&AS, 111, 115



\bibitem[\protect\citeauthoryear{Yoon \& Im}{2020}]{Yoon20} Yoon, Y., Im, M., 2020, ApJ, 893, 117
\bibitem[\protect\citeauthoryear{Yoshino \& Yamauchi}{2015}]{Yoshino15} Yoshino, A., Yamauchi, C., 2015, MNRAS, 446, 3749
\end{thebibliography}
\end{document}